\documentclass{jfm}

\usepackage{graphicx}
\usepackage{amssymb}
\usepackage{color}
\usepackage{float}
\usepackage{verbatim}
\usepackage{amsmath}
\usepackage{wasysym}
\usepackage{tabularx}
\usepackage{natbib}
\usepackage{subfigure}

\definecolor{mygreen}{rgb}{0,0.5,0}

\begin{document}
\title[The effect of phase change on stability of convective flow]{The effect of phase change on stability of convective flow in a layer of volatile liquid driven by a horizontal temperature gradient}

\author{Roman O. Grigoriev, Tongran Qin}
\affiliation{School of Physics, Georgia Institute of Technology, Atlanta, GA 30332-0430, USA}

\maketitle

\begin{abstract}
{Buoyancy-thermocapillary convection in a layer of volatile liquid driven by a horizontal temperature gradient arises in a variety of situations}. 
Recent studies have shown that the composition of the gas phase, which is typically a mixture of vapour and air, has a noticeable effect on the critical Marangoni number describing the onset of convection as well as on the observed convection pattern. 
Specifically, as the total pressure or, equivalently, the average concentration of air is decreased, the threshold of the instability leading to the emergence of convective rolls is found to increase rather significantly. 
We present a linear stability analysis of the problem which shows that this trend can be readily understood by considering the transport of heat and vapour through the gas phase. 
In particular, we show that transport in the gas phase has a noticeable effect even at atmospheric conditions, when phase change is greatly suppressed.
\end{abstract}

\section{Introduction}
\label{sec:intro}

Convection in fluids with a free surface driven by a horizontal temperature gradient has been studied extensively due to applications in crystal growth and thermal management. 
The first systematic study of convection in nonvolatile fluids appears to be due to \citet{Birikh1966} who derived an analytic solution for a planar return flow in a {uniform} laterally unbounded layer due to {a combination of} buoyancy and thermocapillary stresses. 
This solution also describes the flow away from the end walls in a laterally bounded geometry: thermocapillary stresses drive the flow from the hot end towards the cold end near the free surface, with a return flow near the bottom of the layer. 
\citet{kirdyashkin1984} repeated Birikh's theoretical analysis and validated the analytical solutions experimentally.

\citet{Smith1983a,Smith1983b} performed a linear stability analysis of such flows in the limit of vanishing dynamic Bond number, $Bo_D$ (i.e., ignoring buoyancy effects). 
They predicted that, depending on the Prandtl number of the liquid, the uniform base flow would undergo an instability towards either surface waves (for $Pr<0.15$, which corresponds to liquid metals) or hydrothermal waves (for $Pr>0.15$, which corresponds to gases and nonmetallic liquids) above a critical Marangoni number $Ma$, which characterizes the magnitude of thermocapillary stresses. 
In particular, hydrothermal waves were predicted to form at an angle to the direction of the thermal gradient and travel in the direction of thermal gradient. 
As $Pr$ increases, the angle changes smoothly from nearly transverse to nearly parallel to the thermal gradient. 
The theoretical predictions have since been thoroughly tested and verified both in microgravity and for thin films in terrestrial conditions. 
A thorough overview of these experiments is presented in a review paper by \citet{Schatz2001}.

A different type of instability is found at $Bo_D=O(1)$, when buoyancy is nonnegligible. \citet{Villers1992} studied buoyancy-thermocapillary convection in a rectangular cavity for acetone ($Pr=4.24$) experimentally and numerically. 
Although acetone is volatile, reasonable agreement was found between the experimental observations at atmospheric conditions and the numerical simulations based on a one-sided model that ignored heat and mass transfer in the gas phase. 
For low $Ma$ a featureless planar return flow was found, which is well-described by Birikh's solution. 
At higher $Ma$ a steady cellular pattern featuring multiple convection rolls was found instead of hydrothermal waves. 
The (transverse) convection rolls were found to rotate in the same direction, unlike the case of pure buoyancy (or Rayleigh-B\'enard) convection driven by a vertical temperature gradient. 
At even higher $Ma$ the steady state was found to be replaced by an oscillatory pattern that was also unlike a hydrothermal wave: the convection rolls were observed to travel in the direction opposite to that of the thermal gradient.
 Similar results were obtained later by \citet{Desaedeleer1996} for decane ($Pr=15$) and \citet{Garcimartin1997} for 0.65 cSt and 2.0 cSt silicone oil ($Pr=10$  and 30, respectively) in rectangular cavities with strong confinement in the spanwise direction. 

\citet{riley1998} performed one of the most extensive and detailed experimental studies of convection in 1 cSt silicone oil ($Pr=13.9$) in a rectangular cavity with a spanwise dimension {$L_y$} comparable to the streamwise dimension {$L_x$}. 
They discovered that a direct transitions from steady, unicellular flow to hydrothermal waves takes place for small values of the dynamic Bond number ($Bo_D\lesssim 0.2$), while for $Bo_D\gtrsim 0.2$ the results are similar to those of earlier studies with spanwise confinement: the featureless return flow first transitions to steady co-rotating convection cells and, upon further increase in $Ma$, to an oscillatory multicellular pattern. 
Riley and Neitzel also determined the critical values of $Ma$ and the wavelength $\lambda$ of the convective pattern as a function of $Bo_D$. 
\citet{Burguete2001} performed experiments on convection in 0.65 cSt silicone oil ($Pr=10.3$) in a rectangular cavity with different aspect ratios where spanwise dimension was greater than the streamwise dimension. 
Similarly, they found that the base return flow destabilizes into either oblique travelling waves or longitudinal stationary rolls, respectively, for low and high $Bo_D$.

Convective patterns have also been studied extensively using numerical simulations. 
Most of the numerical studies \citep{BenHadid1990,Villers1992,BenHadid1992,Mundrane1994,Lu1998,Shevtsova2003} were based on one-sided models which ignore the transport in the gas phase, assume that the free surface is flat and nondeformable, the bottom wall and the interface are adiabatic, and phase change is negligible. 
These numerical simulations were able to reproduce some features of the experimental studies. 
For example, \citet{Shevtsova2003} and \citet{Shevtsova2003b} performed numerical simulations for decane ($Pr = 14.8$) in a rectangular layer at different $Bo_D$. 
They found that as $Ma$ increases, the primary instability leads to hydrothermal waves for $Bo_D \le 0.25$, while for $Bo_D \ge 0.32$ the primary (secondary) instability produces a steady (oscillatory) multicellular flow.

Since it does not account for buoyancy, the linear stability analysis of \citet{Smith1983a,Smith1983b} fails to predict the stationary patterns that emerge for $Bo_D=O(1)$. 
However, most of the linear stability analyses accounting for buoyancy also failed to predict the correct pattern, i.e., stationary (transverse) {rolls} that were observed in the experiments. 
Using adiabatic boundary conditions at the top and bottom of the liquid layer, \citet{Parmentier1993} predicted transition to travelling waves rather than steady multicellular pattern for a range of $Pr$ from 0.01 to 10, regardless of the value of $Bo_D$. \citet{Chan2010}, who used similar assumptions, also predicted transition to travelling waves for a $Pr=13.9$ fluid. 
Moreover, their predicted critical $Ma$ and wavelength $\lambda$ do not match the experiment \citep{riley1998}. 
In both cases the predicted travelling waves are oblique for smaller $Bo_D$ and become transverse for $Bo_D$ greater than some critical $O(1)$ value.

\citet{Mercier1996} showed that transition to a stationary convective pattern can take place if the adiabatic boundary conditions are replaced with Newton's cooling law, although that requires an unrealistically large surface Biot number ($Bi\gtrsim 185/Bo_D$). 
Moreover, the predicted pattern corresponds to longitudinal convection rolls, while in most experiments transverse rolls were observed. 
In a subsequent paper \citet{Mercier2002} considered the effects of the end walls, which they described as spatial disturbances superimposed on the uniform base flow. 
Their analysis predicted that, depending upon the Prandtl number, convection rolls would develop near the hot end (for $Pr>4$), near the cold end (for $Pr<0.01$), or at both end walls (for $0.01<Pr<4$).

To our knowledge, the study by \citet{Priede1997} is the only one to date which correctly predicts the formation of a stationary pattern at $Bo_D=O(1)$. 
They argued that travelling waves, being convectively unstable, cannot get sufficiently amplified by linear instability in a laterally bounded system. 
At the same time, the {effect of lateral walls extends far into the bulk of the liquid layer, resulting in a stationary pattern of transverse convection rolls}. 
Their predicted critical values of $Ma$ are in reasonable agreement with the threshold values found by \citet{riley1998}, although {their is a systematic deviation}, suggesting that some important effects have not been considered.

The volatility of the fluids used in the experiments is one source of the discrepancy with analytical (and most numerical) predictions. 
Although at atmospheric conditions phase change is usually strongly suppressed, it can still play a role. 
The latent heat associated with phase change can significantly modify the interfacial temperature, and hence the thermocapillary stresses. 
However, there are very few studies that investigated this effect. \citet{YouRong2012} have studied nonadiabatic effects by using Newton's cooling law with a small Biot number. 
Their numerical simulations ignored phase change, but were able to reproduce many features of the experimental observations at atmospheric conditions. \citet{Ji2007} considered phase change, but ignored buoyancy, so their analysis is only applicable for thin films or under microgravity, i.e., when $Bo_D\approx 0$.

A few recent studies investigated the role played by the gas phase, which is generally a binary mixture of a noncondensable gas (typically air) and the vapour, in more detail. 
In particular, the experimental study of \citet{Li2013}, which used a volatile 0.65 cSt silicone oil ($Pr=9.2$), showed that for {$Bo_D=O(1)$} the threshold values of $Ma$ increase rather dramatically as the air is removed from the experimental apparatus. 
The numerical simulations of this experimental setup \citep{Qin2013,Qin2014,Qin2015} based on a two-sided model, which took phase change into account and explicitly described the transport of heat and vapour through the gas phase, reproduced the experimental results.  
The main effect of air is to suppress phase change by impeding the transport of vapour towards, or away from, the interface, but its presence also affects thermal conductivity of the gas layer. 
This suggests that phase change and transport in the gas phase play an important role in this problem, especially at reduced pressures (reduced concentrations of air).

In order to better understand the mechanism of the instability and the effect of the gas phase we {formulated a two-sided model of the flow valid for arbitrary composition of the gas phase; it is} described in Section \ref{sec:model}. 
The analytical solution describing the uniform return flow (in both phases) is derived in Section \ref{sec:suf}. 
The stability of that solution is investigated in Section \ref{sec:smf}. 
The results of linear stability analysis are compared with {previous analytical}, experimental, and numerical studies in Section \ref{sec:res} and are discussed in Section \ref{sec:dis}. 
Conclusions are presented in Section \ref{sec:sum}.

\section{Mathematical Model}
\label{sec:model}

\subsection{Governing Equations}

We will assume that the liquid layer has depth $d_l$ and lateral dimensions much larger than $d_l$. 
Since in most experiments the system is covered by a horizontal plate to limit evaporation, we will assume that the gas layer also has a finite depth $d_g$.  
To describe convection in this two-layer system, we will use a variation of the two-sided model originally introduced by \citet{Qin2013} for near-atmospheric conditions and later extended by \citet{Qin2015} to the limit when the gas phase is dominated by the vapour, rather than air. 
A version of the model {interpolating between the two} limits is summarized below. 
The momentum transport in the bulk is described, for both the liquid and the gas phase, by the Navier-Stokes equation
\begin{equation}
\label{eq:ns}
\rho \left({\partial }_t{\bf u}+{\bf u}\cdot \nabla {\bf u}\right)=-\nabla p+\mu {\nabla }^2{\bf u}+\rho \left(T,c_a\right){\bf g}
\end{equation}
where $p$ is the fluid pressure, $\rho$ and $\mu$ are the fluid's density and viscosity, respectively, $c_a$ is the concentration of air, and ${\bf g}$ is the gravitational acceleration. (The air is noncondensable, so $c_a=0$ in the liquid phase.)
{Following standard practice, we use the Boussinesq approximation, where the density is considered constant everywhere except in the last term on the right-hand-side representing the buoyancy force. 
In particular, the mass conservation equation in both phases reduces to
\begin{equation}
\label{eq:divu}
\nabla\cdot{\bf u}=0.
\end{equation}
To account for buoyancy, the density of the liquid phase is assumed to depend linearly on the temperature}
\begin{equation}
\label{eq:rhol}
\rho_l=\rho^0_l[1-\beta_l\left(T-T_0\right)],
\end{equation}
where {$\rho^0_l$ is the density at the temperature $T_0$ describing global thermodynamic equilibrium}, and $\beta_l=-\rho_l^{-1}\partial \rho_l/\partial T$ is the coefficient of thermal expansion. 
Here and below, subscripts $l$, $g$, $v$, $a$, and $i$ denote properties of the liquid and gas phase, vapour and air component, and the liquid-vapour interface, respectively. {The index $0$ will be used throughout the paper to distinguish the equilibrium values from the nonequilibrium ones, where the choice is not obvious.} 
In the gas phase
\begin{equation}
\label{eq:rhog}
\rho_g=\rho_a+\rho_v,
\end{equation}
where both vapour and air are considered to be ideal gases
\begin{equation}
\label{eq:ideal-gas}
\rho_{a,v}=\frac{p_{a,v}}{{\bar{R}}_{a,v}T},
\end{equation}
${\bar{R}}_{a,v}=R/M_{a,v}$, $R$ is the universal gas constant, and $M_{a,v}$ is the molar mass of air/vapour.
{Correspondingly, the total pressure in the gas is}
\begin{equation}
\label{eq-7}
p_g=p_a+p_v.
\end{equation}

{Mass transport in the gas phase is described by the standard conservation equation(s)
\begin{align}
\label{eq:adv-n}
{\partial }_t n_{a,v}+\nabla\cdot{\bf j}_{a,v}=0,
\end{align}
for the number density of the two components
\begin{align}
n_{a,v}=\frac{\rho_{a,v}}{m_{a,v}},
\end{align}
where $m_{a,v}=M_{a,v}/N_A$ is the mass of one air/vapour molecule.
The number density flux is given by
\begin{align}
\label{eq:j}
{\bf j}_{a,v}= n_{a,v}{\bf u}-n_gD{\nabla }c_{a,v}=n_g({\bf u}c_{a,v}-D{\nabla }c_{a,v}),
\end{align}
where the first and the second term on the right-hand-side represent the contributions due to advection and diffusion, respectively, $D$ is the binary diffusion coefficient, and
\begin{equation}
\label{eq-9}
c_{a,v}=\frac{n_{a,v}}{n_g}=\frac{p_{a,v}}{p_g}
\end{equation}
are the concentrations (or, more precisely, the molar fractions) of the two components. 
Therefore, the conservation equation(s) \eqref{eq:adv-n} can be rewritten as
\begin{align}
\label{eq:adv-nv}
\partial_t (n_gc_{a,v})+n_g{\bf u}\cdot\nabla c_{a,v}=\nabla \cdot ({n_g}D\nabla c_{a,v}).
\end{align}
From the ideal gas law, the total number density in the gas phase
\begin{equation}
\label{eq:rhox}
n_g=\frac{p_g}{k_BT},
\end{equation}
where $k_B=R/N_A$ is the Boltzman constant. 
As we showed previously \citep{Qin2014}, spatial variation of $p_g$ can in practice be neglected. 
Furthermore, the largest temperature variation in relevant experimental and numerical studies is around $5\%$ (and typically much smaller than that), such that $n_g$ can be assumed constant. Consequently \eqref{eq:adv-nv} reduces to an advection-diffusion equation for the two concentrations
\begin{align}
\label{eq:adv-cv}
\partial_t c_{a,v}+{\bf u}\cdot\nabla c_{a,v}=\nabla \cdot (D\nabla c_{a,v}).
\end{align}
These two equations are equivalent, so either one can be used, since $c_a+c_v=1$.}

Finally, the transport of heat is also described using an advection-diffusion equation
\begin{equation}
\label{eq:adv-T}
{\partial }_tT+{\bf u}\cdot\nabla T=\alpha \nabla^2T,
\end{equation}
where $\alpha=k/(\rho C_p)$ is the thermal diffusivity, $k$ is the thermal conductivity, and $C_p$ is the heat capacity, of the fluid.

\subsection{Boundary Conditions at the Interface}

The system of coupled evolution equations for the velocity, pressure, temperature, and {concentration} fields should be solved in a self-consistent manner, subject to the boundary conditions describing the balance of momentum, heat, and mass fluxes {between the two phases}. 
The phase change at the interface can be described using Kinetic Theory \citep{schrage1953}. 
As we have shown previously \citep{Qin2014}, the choice of the phase change model has negligible effect on the results. 
The mass flux across the interface is given by \citep{ajaev2009}
\begin{equation}
\label{eq:ktg}
J={\frac{2\chi}{2-\chi}}\,\rho_v\sqrt{\frac{\bar{R}_vT_i}{2\pi }}\left[\frac{p_l-p_g}{\rho_l\bar{R}_vT_i}+\frac{\mathcal{L}}{\bar{R}_vT_i}\frac{T_i-T_{s}}{T_{s}}\right],
\end{equation}
where $\chi$ is the accommodation coefficient, $\mathcal{L}$ is the latent heat of vaporization, and subscript $s$ denotes saturation values for the vapour. 
%
{The dependence of the local saturation temperature $T_s$ on the partial pressure of vapour $p_v$ is described using the Clausius-Clapeyron equation for phase equilibrium
\begin{equation}
\label{eq:CC}
\ln \frac{p_v}{p_v^0}=\frac{\mathcal{L}}{\bar{R}_v}\left[\frac{1}{T_0}-\frac{1}{T_s}\right],
\end{equation}
where $p_v^0$ is the saturation pressure of the vapour at the equilibrium temperature $T_0$.}
The first term in \eqref{eq:ktg} is proportional to the Young-Laplace pressure and can be ignored in this problem, since the interface is {considered} flat. 

The mass flux balance at the interface can be expressed with the help of (\ref{eq:adv-n}). In the reference frame of the interface, the mass flux of the vapour is given by
\begin{equation}
\label{eq:vap-flux}
\frac{J}{m_v}={\hat{\bf n}}\cdot{\bf j}_v=n_g\hat{\bf n}\cdot([{\bf u}_g - {\bf u}_i] c_v
-D\nabla c_v),
\end{equation}
where ${\bf u}_i$ is the velocity of the liquid at the interface. Since air is noncondensable, its mass flux across the interface is zero:
\begin{equation}
\label{eq:air-flux}
0={\hat{\bf n}}\cdot{\bf j}_a=n_g\hat{\bf n}\cdot([{\bf u}_g - {\bf u}_i] c_a
-D\nabla c_a).
\end{equation}
Since $c_a+c_v=1$, these two relations can be solved yielding two of the boundary conditions for \eqref{eq:ns} and \eqref{eq:adv-n} in the gas phase
\begin{equation}\label{eq:ns4}
\hat{\bf n}\cdot\nabla c_v=-\frac{c_aJ}{m_vn_gD}
\end{equation}
and
\begin{equation}\label{eq:ns5}
\hat{\bf n}\cdot({\bf u}_g - {\bf u}_i) =\frac{J}{m_vn_g}.
\end{equation}

The heat flux balance is given by
\begin{equation}
\label{eq:q-balance}
\mathcal{L}J=\hat{\bf n}\cdot k_g\nabla T_g-\hat{\bf n}\cdot k_l\nabla T_l,
\end{equation}
where the advective contribution to the heat flux is negligible on both sides of the interface. Indeed, in the gas phase, conduction is the dominant contribution {\citep{Qin2015}}, while on the liquid side
\begin{equation}
\label{eq:ns8}
\hat{\bf n}\cdot \left({\bf u}_l-{\bf u}_i\right)=\frac{J}{\rho_l}.
\end{equation}
Since the liquid density is much greater than that of the gas, the left-hand-side of (\ref{eq:ns8}) is very small compared with $\hat{\bf n}\cdot({\bf u}_g-{\bf u}_i)$ and can be ignored.

The remaining boundary conditions for $\bf u$ and $T$ at the liquid-vapour interface are standard: the temperature is continuous
\begin{equation}\label{eq:bct}
T_l=T_i=T_v
\end{equation}
as are the tangential velocity components
\begin{equation}\label{eq:bcvt}
{(\mathbb{I}- \hat{\bf n} \hat{\bf n}) \cdot (\bf u_l - \bf u_g)= 0.
}
\end{equation}
The stress balance
\begin{equation}
\label{eq:stress}
({\Sigma }_l-{\Sigma }_g{\rm )}\cdot \hat{\bf n}
= \hat{\bf n}\kappa{\sigma }-{\gamma }{\nabla }_sT_i
\end{equation}
incorporates both the viscous drag between the two phases and the thermocapillary effect. Here $\Sigma =\mu \left[\nabla {\bf u}+{(\nabla {\bf u})}^T\right]-p$ is the stress tensor, $\kappa${$(=0)$} is the interfacial curvature, 
{${\nabla }_s=\left(\mathbb{I}-{\hat{\bf n}} \hat{\bf n}\right)\cdot\nabla$
}
is the surface gradient, and $\gamma=-\partial\sigma/\partial T$ is the temperature coefficient of surface tension.

{We will further assume that side/top/bottom walls are adiabatic, $\hat{\bf n}\cdot\nabla T=0$, which is a good approximation for most experimental setups.
The heat transport through the end walls at $x=0$ and $x=L_x$ is not treated explicitly in this study, but in principle a variety of boundary conditions, from constant temperature to constant flux to mixed boundary conditions can be accommodated.  
Standard no-slip boundary conditions ${\bf u}=0$ for the velocity and no-flux boundary conditions
$\hat{\bf n}\cdot\nabla c_{a,v}=0$ for concentration apply on all the walls. }

\section{The Base Flow}
\label{sec:suf}

In liquid layers that are not too thin, under normal gravity, convection is driven by both buoyancy and thermocapillarity. 
{The strength of these two effects} can be quantified, respectively, in terms of the Marangoni number
\begin{equation}\label{eq:Mai}
Ma=\frac{\gamma d_l^2}{\mu_l\alpha_l}\tau
\end{equation}
and the Rayleigh number
\begin{equation}\label{eq:Ra}
Ra=\frac{g\beta_ld_l^4\tau}{\nu_l\alpha_l},
\end{equation}
where $\nu_l=\mu_l/\rho_l$ is the kinematic viscosity of the liquid and $\tau$ is the {horizontal component of the temperature gradient, assumed to be in the positive $x$ direction. In the numerics and experiment we set $\tau=\hat{\bf x}\cdot\nabla_s T_i$, which is found to be nearly independent of the location in the central region of the flow \citep{Qin2015}}. The dynamic Bond number
\begin{equation}\label{eq:BoD}
Bo_D=\frac{Ra}{Ma}=\frac{\rho_lg\beta_ld_l^2}{\gamma}
\end{equation}
is independent of $\tau$ and quantifies the relative strength of buoyancy and thermocapillarity. 
In defining nondimensional combinations, as well as various scales, we will use the values of {material parameters at the equilibrium temperature $T_0$.}

\begin{figure}
\subfigure[]{\includegraphics[width=.49\columnwidth]{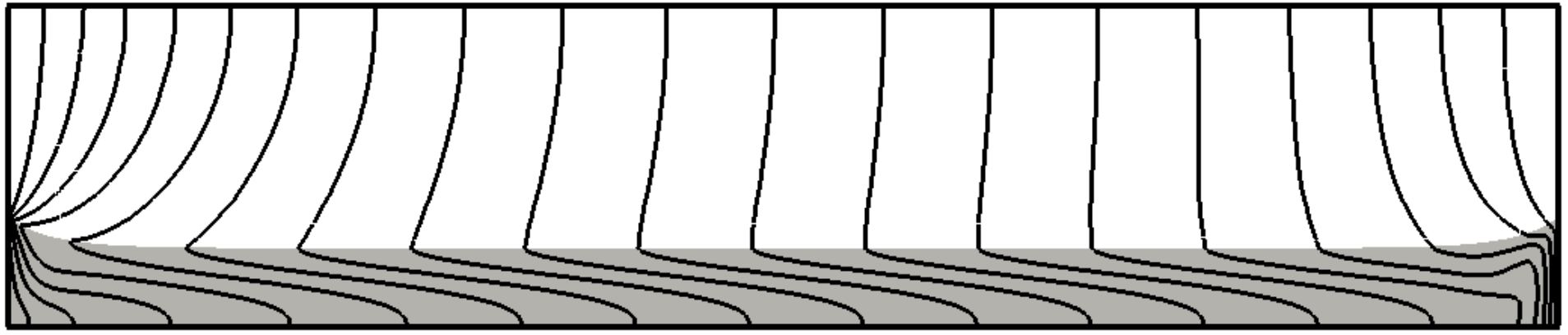}}
\subfigure[]{\includegraphics[width=.49\columnwidth]{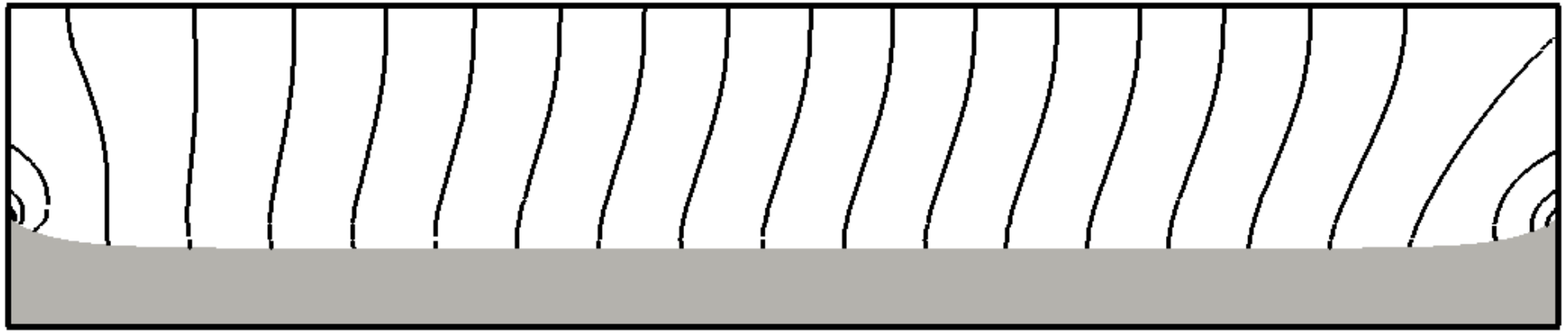}}\\
\subfigure[]{\includegraphics[width=.49\columnwidth]{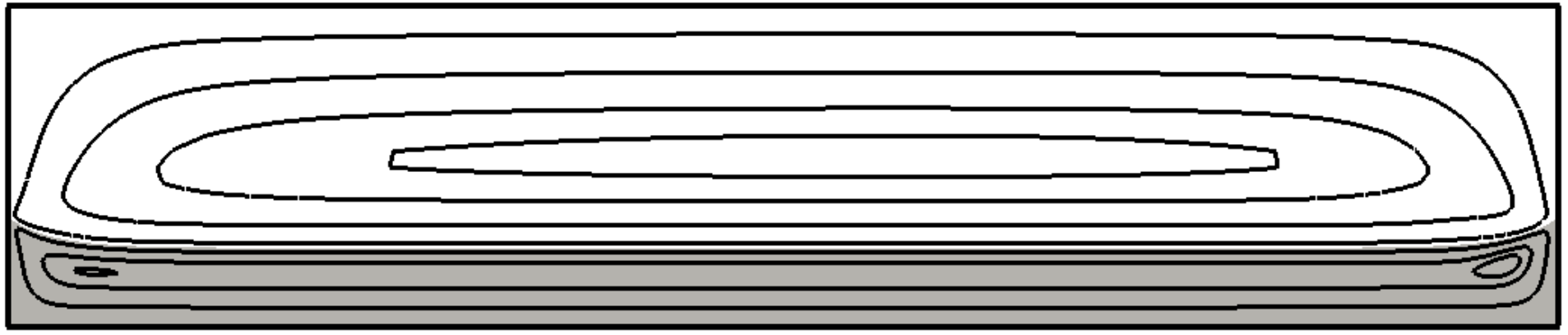}}
\subfigure[]{\includegraphics[width=.49\columnwidth]{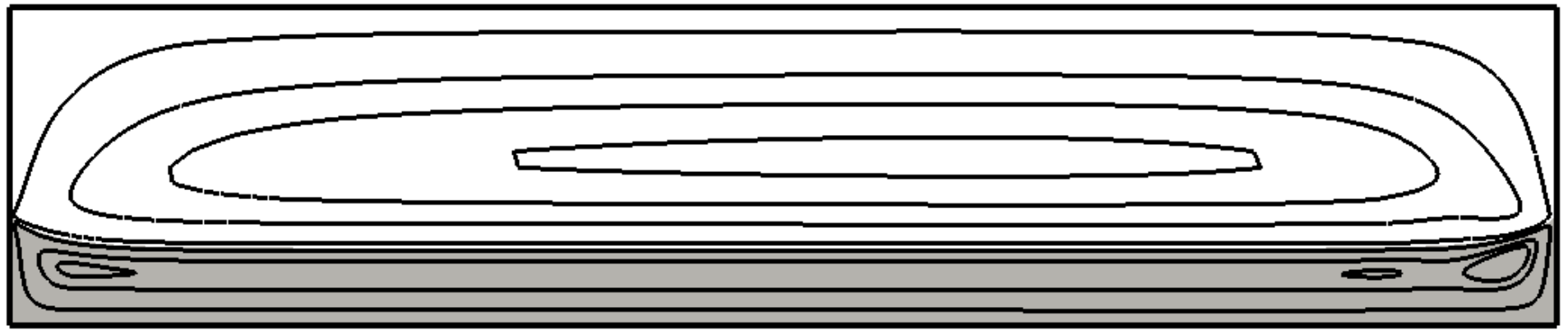}}
\caption{{Numerical solutions at $c_a^0 = 0.96$ (atmospheric conditions) for the flow of 0.65 cSt silicone oil in a rectangular cell with aspect ratios $\Gamma_x=19.4$ and $A=3$ \citep{Qin2015}. Shown are the level sets of (a) the temperature field $T$, (b) vapour concentration field $c_v$, and (c) stream function $\psi$ for $Ma=190$. Panel (d) shows the streamlines for a slightly higher $Ma=370$. Here and below, the gray (white) background indicates the liquid (gas) phase. The horizontal and vertical axes are $x$ and $z$; the cold/hot end wall is on the left/right. }
}
\label{fig:SUF}
\end{figure}

{As discussed in the introduction, at} sufficiently low $Ma$, a steady return flow is found in the liquid layer.
{The analytical solution of \citet{Birikh1966} describes such a flow in laterally unbounded layers of nonvolatile liquids.}
This solution also describes the flow observed in laterally bounded layers away from the end walls {(cf. Fig. \ref{fig:SUF}) even for volatile fluids under atmospheric conditions when the phase change is strongly suppressed \citep{Qin2013}.}
For volatile fluids at reduced pressures, Birikh's solution becomes invalid due to the increasing role of phase change  \citep{Li2013,Qin2013}. 
Instead, we should look for a solution to the two-sided model described in the previous section, which describes the flow in both layers. 
In order to reduce the number of parameters, the governing equations \eqref{eq:ns}, \eqref{eq:divu}, \eqref{eq:adv-cv}, and \eqref{eq:adv-T}  are nondimensionalized by introducing the length scale $d_l$, time scale $d_l^2/\nu_l$, velocity scale $\nu_l/d_l$, density scale $\rho_l$, pressure scale $\rho_l(\nu_l/d_l)^2$, and temperature scale $\tau d_l=\mu_l\alpha_lMa/(\gamma d_l)$.

The dimensionless governing equations for the liquid layer become
\begin{align}
\tilde{\nabla} \cdot \tilde{\bf u}_l&=0,\nonumber\\
\partial_{\tilde{t}}\tilde{\bf u}_l+\tilde{\bf u}_l\cdot \tilde{\nabla}\tilde{\bf u}_l &= -\tilde{\nabla}\tilde{p}+\tilde{\nabla}^2\tilde{\bf u}_l+Gr\tilde{T}_l\hat{z},\nonumber\\
\label{eq:geq3}
\partial_{\tilde{t}}\tilde{T}_l+\tilde{\bf u}_l\cdot\tilde{\nabla}\tilde{T}_l&=Pr^{-1} \tilde{\nabla} ^2\tilde{T}_l,
\end{align}
where $\tilde{\nabla}=d_l\nabla$, $\tilde{\bf u}_l={\bf u}_ld_l/\nu_l$, $\tilde{T}_l=(T_l-T_0)/(\tau d_l)$, and  
\begin{align}\label{eq:Gr}
Gr=\frac{Ra}{Pr}=\frac{g\beta_ld_l^4\tau}{\nu_l^2}
\end{align}
is the Grashof number. We will use a coordinate system defined such that the liquid-vapour interface corresponds to the plane $\tilde{z}=0$ (so that the liquid layer corresponds to $-1<\tilde{z}<0$). {Recall that the $x$ axis points in the direction of the applied temperature gradient, with the origin chosen such that $T_i=T_0$ at $\tilde{x}=0$.}

The dimensionless governing equations for the gas layer ($0<\tilde{z}<A$) are
\begin{align}
\tilde{\nabla} \cdot \tilde{\bf u}_g&=0,\nonumber\\
\partial_{\tilde{t}}\tilde{\bf u}_g+\tilde{\bf u}_g\cdot \tilde{\nabla}\tilde{\bf u}_g &= -\frac{\rho_l}{\rho_g}\tilde{\nabla}\tilde{p}+K_\nu\tilde{\nabla}^2\tilde{\bf u}_g+(\Xi_T\tilde{T}_g+\Xi_c\tilde{c}_v)\hat{z},\nonumber\\
\partial_{\tilde{t}}\tilde{T}_g+\tilde{\bf u}_g\cdot\tilde{\nabla}\tilde{T}_g&=K_\alpha\tilde{\nabla} ^2\tilde{T}_g,\nonumber\\
\label{eq:geq4}
\partial_{\tilde{t}} \tilde{c}_v+\tilde{\bf u}_g\cdot\tilde{\nabla}\tilde{c}_v&=K_D\tilde{\nabla}^2\tilde{c}_v,
\end{align}
where $\tilde{\bf u}_g={\bf u}_gd_l/\nu_l$, $\tilde{T}_g=(T_g-T_0)/(\tau d_l)$, $\tilde{c}_v=c_v-c_v^0$,  $c_v^0=1-c_a^0$,
$A=d_g/d_l$, 
$K_\nu={\nu_g}/{\nu_l}$, $K_\alpha={\alpha_g}/{\nu_l}$, and $K_D=D/\nu_l$.
The nondimensional combinations
\begin{align}\label{eq:Xi}
\Xi_T&=
\frac{g\beta_gd_l^4\tau}{\nu_l^2}=\frac{\beta_g}{\beta_l}Gr,\nonumber\\
\Xi_c&=\frac{gd_l^3}{\nu_l^2}\frac{m_a-m_v}{c_a^0m_a+c_v^0m_v}
\end{align}
describe the contributions to the buoyancy force in the gas layer due to perturbations in the temperature and composition of the gas, respectively. Note that $\Xi_T$ depends on the imposed temperature gradient $\tau$, but not $c_a^0$, while $\Xi_c$ depends on $c_a^0$, but not $\tau$.

Both the imposed temperature gradient $\tau$ and the composition of the gas phase, parametrized by the equilibrium concentration of air $c_a^0$, play a key role in this problem. In experiment \citep{Li2013}, $c_a^0$ was controlled indirectly by varying the net gas pressure
\begin{align}
\label{eq:pg0}
p_g^0=\frac{p_v^0}{1-c_a^0}.
\end{align}
In the following analysis, it will be more convenient to describe the composition directly in terms of $c_a^0$. 
{In terms of $p_g^0$, we can write $n_g\approx p_g^0/(k_BT_0)$, while $\tilde{p}$ in \eqref{eq:geq3} and \eqref{eq:geq4} represents the nondimensional form of the difference $p_g-p_g^0$.}

In order to satisfy the incompressibility condition, we will assume that the flow is strictly two-dimensional and introduce a stream function for each layer, such that
\begin{align}
\tilde{\bf u}_l&=(\partial_{\tilde{z}}\tilde{\psi}_l,0,-\partial_{\tilde{x}}\tilde{\psi}_l),\nonumber\\
\tilde{\bf u}_g&=(\partial_{\tilde{z}}\tilde{\psi}_g,0,-\partial_{\tilde{x}}\tilde{\psi}_g).
\end{align}
Eliminating the pressure, the governing equations (\ref{eq:geq3}) for the liquid layer can be rewritten as
\begin{align}
(\partial_{\tilde{t}} - \tilde{\nabla}^2 
+\partial_{\tilde{z}}\tilde{\psi}_l\partial_{\tilde{x}}
- \partial_{\tilde{x}}\tilde{\psi}_l\partial_{\tilde{z}})\tilde{\nabla}^2\tilde{\psi}_l 
+ Gr\partial_{\tilde{x}}\tilde{T}_l &=0,\nonumber\\
\label{eq:nliq}
{\partial_{\tilde{t}}}\tilde{T}_l+{\partial_{\tilde{z}} \tilde{\psi}_l}{\partial_{\tilde{x}}}\tilde{T}_l - {\partial_{\tilde{x}} \tilde{\psi}_l}{\partial_{\tilde{z}}}\tilde{T}_l -Pr^{-1}\tilde{\nabla}^2\tilde{T}_l&=0.
\end{align}
For the gas layer we have
\begin{align}
(\partial_{\tilde{t}} -K_\nu\tilde{\nabla}^2
+\partial_{\tilde{z}}\tilde{\psi}_g\partial_{\tilde{x}}
- \partial_{\tilde{x}}\tilde{\psi}_g\partial_{\tilde{z}})\tilde{\nabla}^2\tilde{\psi}_g
+ \Xi_T\partial_{\tilde{x}}\tilde{T}_g
+ \Xi_c\partial_{\tilde{x}}\tilde{c}_v &= 0,\nonumber\\
{\partial_{\tilde{t}}}\tilde{T}_g+{\partial_{\tilde{z}} \tilde{\psi}_g}{\partial_{\tilde{x}}}\tilde{T}_g 
- {\partial_{\tilde{x}} \tilde{\psi}_g}{\partial_{\tilde{z}}}\tilde{T}_g-K_\alpha\tilde{\nabla}^2\tilde{T}_g&=0,\nonumber\\
\label{eq:ngas}
{\partial_{\tilde{t}}}\tilde{c}_v+{\partial_{\tilde{z}} \tilde{\psi}_g}
{\partial_{\tilde{x}}}\tilde{c}_v-{\partial_{\tilde{x}} \tilde{\psi}_g}{\partial_{\tilde{z}}}\tilde{c}_v
-K_D\tilde{\nabla}^2\tilde{c}_v&=0.
\end{align}

\subsection{Boundary Conditions}

At the bottom of the liquid layer ($\tilde{z}=-1$) and the top of the gas layer ($\tilde{z}=A$), no-slip and adiabatic boundary conditions apply
\begin{align}
\label{eq:bc1}
\tilde{\bf u}&=0,\nonumber\\
\partial_{\tilde{z}} \tilde{T} &= 0.
\end{align}
At the interface ($\tilde{z} = 0$), the temperature and velocity fields are continuous
\begin{align}\label{eq:bc2}
\tilde{T}_l &= \tilde{T}_g = \tilde{T}_i,\nonumber\\
\tilde{\bf u}_l &= \tilde{\bf u}_g = \tilde{\bf u}_i,
\end{align}
where {for our choice of the origin of coordinate system} $\tilde{T}_i=\tilde{x}$.
Since $\mu_l\gg\mu_g$ and typically $d_g>d_l$, the viscous stress in the gas layer can be ignored, yielding a simplified expression for the shear stress balance at the interface
\begin{align}
\label{eq:ssb}
\partial_{\tilde{z}} \tilde{u}_{l,x} &= -Re \partial_{\tilde{x}} \tilde{T}_l,
\end{align}
where
\begin{align}\label{eq:Re}
Re=\frac{Ma}{Pr}=\frac{\gamma d_l^2}{\mu_l\nu_l}\tau,  
\end{align}
is the Reynolds number.

The heat flux balance (\ref{eq:q-balance}) at the interface reduces to
\begin{align}
\label{eq:bc4}
\partial_{\tilde{z}} \tilde{T}_l&= \frac{k_g}{k_l} \partial_{\tilde{z}} \tilde{T}_g-\frac{V}{Ma} \tilde{J},
\end{align}
where $k_g = c_v^0k_v + c_a^0 k_a$, and $\tilde{J}=Jd_l/(Dm_vn_g)$ is the dimensionless mass flux. The dimensionless combination 
\begin{align}
\label{eq:Gv}
V=\frac{{\mathcal L}\gamma d_l}{\alpha_l\mu_lk_l}\frac{Dp_g^0}{\bar{R}_vT_0},
\end{align}
or more precisely the ratio $V/Ma$, describes the relative magnitude of the latent heat released (absorbed) at the interface due to condensation (evaporation) compared with the vertical heat flux in the liquid layer due to conduction. It should be noted that, although the product $Dp_g^0$, and consequently $V$, is a function of $T_0$, it is independent of the gas pressure, and therefore $c_a^0$.

Nondimensionalizing \eqref{eq:ns4} and \eqref{eq:ns5}, we obtain
\begin{align}
\label{eq:jcv}
\partial_{\tilde{z}}\tilde{c}_v=-c_a^0c_v^0\tilde{J}
\end{align}
and
\begin{equation}
\label{eq:jun}
\hat{\bf n}\cdot(\tilde{\bf u}_g - \tilde{\bf u}_i) =c_v^0K_D\tilde{J}.
\end{equation}
The mass flux vanishes at the top of the gas layer
\begin{align}
\label{eq:bctop}
\partial_{\tilde{z}}\tilde{c}_v &= 0.
\end{align}
The base uniform flow corresponds to a vanishing mass flux at the interface, $\tilde{J}=0$, which is also an assumption made by all one-sided models. {The base flow is spatially uniform in the lateral direction(s) which, coupled with incompressibility, requires} ${\hat{\bf n}}\cdot{\bf u}_i=0$. This leads to a number of simplifications. In particular, (\ref{eq:jcv}) and (\ref{eq:jun}) require that $\tilde{u}_{l,z}=\tilde{u}_{g,z}=0$ and $\partial_{\tilde{z}}\tilde{c}_v=0$ at the interface. Furthermore, the heat flux at the interface should also vanish, so that \eqref{eq:q-balance} gives $\partial_{\tilde{z}}\tilde{T}_l=\partial_{\tilde{z}}\tilde{T}_g=0$.

The boundary conditions for the stream function can be easily obtained from those for the velocities. At the bottom of the liquid layer and the top of the gas layer
\begin{align}
\tilde{\psi}&=0,\nonumber\\
\partial_{\tilde{x}} \tilde{\psi} = \partial_{\tilde{z}} \tilde{\psi}&=0. \label{eq:bcvgb}
\end{align}
For a uniform flow, the net flux through any vertical plane vanishes. {Combined with conditions \eqref{eq:bcvt} and \eqref{eq:stress} this requires that at the interface}
\begin{align}
\tilde{\psi}_l=\tilde{\psi}_g&=0,\nonumber\\
\partial_{\tilde{z}}\tilde{\psi}_l-\partial_{\tilde{z}}\tilde{\psi}_g&=0,\nonumber\\
\partial_{\tilde{z}}^2 \tilde{\psi}_l + Re\partial_{\tilde{x}} \tilde{T}_l&=0. \label{eq:bcvgt}
\end{align}

\subsection{Fluid Flow and Temperature in the Liquid Layer}

For a {uniform} horizontal flow, where both $\tilde{\psi}_l$ and $\tilde{\psi}_g$ are functions of $\tilde{z}$ alone, we can look for solutions to \eqref{eq:nliq} in the form 
\begin{align}
\tilde{T}_l = \tilde{x}+\tilde{\theta}_l(\tilde{z}),
\end{align}
where $\tilde{\theta}_l(0)=0$. With this choice, the system (\ref{eq:nliq}) reduces to coupled ordinary differential equations (ODEs)
\begin{align}
-\tilde{\psi}_l'''' + Gr &= 0,\nonumber\\
\label{eq:baseeq2}
Pr\tilde{\psi}_l' - \tilde{\theta}''_l&=0,
\end{align}
where prime stands for the derivatives with respect to the $\tilde{z}$ coordinate.

Solving the system (\ref{eq:baseeq2}) subject to the boundary conditions at the bottom and the free surface of the liquid layer, we find the steady state solutions for the stream function
\begin{equation}
\label{eq:psitheo}
\tilde{\psi}_l = Re\left[-\frac{\tilde{z}(\tilde{z}+1)^2}{4} 
+ Bo_D\frac{\tilde{z}(\tilde{z}+1)^2(2\tilde{z}-1)}{48}\right],
\end{equation}
velocity
\begin{equation}
\label{eq:utheo}
\tilde{\bf u}_l = Re\left[-\frac{(\tilde{z}+1)(3\tilde{z}+1)}{4} 
+ Bo_D\frac{(\tilde{z}+1)(8\tilde{z}^2+\tilde{z}-1)}{48}\right]\hat{x},
\end{equation}
 and temperature field
\begin{equation}
\label{eq:Ttheo}
\tilde{T}_l=\tilde{x}+Ma\left[-\frac{\tilde{z}^2(3\tilde{z}^2+8\tilde{z}+6)}{48} 
+ Bo_D\frac{\tilde{z}^2(8\tilde{z}^3+15\tilde{z}^2-10)}{960}\right]
\end{equation}
describing the base flow. They agree with the analytical solutions originally obtained by Birikh \citep{Birikh1966} and later rederived by Kirdyashkin \citep{kirdyashkin1984} and Villers and Platten \citep{Villers1987} using a one-sided model that ignores the effects of the gas phase.

The assumption that the interfacial temperature varies linearly has been widely used in previous studies, without much justification, for deriving the solutions (\ref{eq:utheo}) and (\ref{eq:Ttheo}) for the return flow underlying the stability analyses \citep{Parmentier1993, Mercier1996, Priede1997} as well as in models of the adiabatic section of heat pipes \citep{Ha1994,Suman2005,Markos2006}, which assume unidirectional flow in the liquid phase. However, the validity of this assumption cannot be established by a one-sided model which ignores {heat and mass} transport in the gas phase. In fact, when $c_a^0$ becomes sufficiently low, the interfacial temperature profile becomes nonlinear \citep{Li2013,Qin2015}. Proper justification of the linearity assumption requires showing that it is consistent with a steady-state solution of the transport equations in the gas phase, which satisfies all of the boundary conditions at the free surface. We turn to this next.

\subsection{Fluid Flow, Temperature, and Composition in the Gas Layer}

The solutions for the velocity, temperature, and composition of the gas phase can be found in the same way the solutions (\ref{eq:utheo}) and (\ref{eq:Ttheo}) were obtained for the liquid phase. We will start by finding the solution for the vapour concentration at the interface before deriving the solution in the bulk. Since there is no phase change, $T_s=T_i$, according to (\ref{eq:ktg}), and $\partial_xT_s=\tau$. Using Clausius-Clapeyron relation  \eqref{eq:CC} and neglecting the deviation of $p_g$ from the equilibrium value \eqref{eq:pg0} we therefore find
\begin{equation}\label{eq:Tsat}
\partial_x c_v=\frac{\mathcal{L}\tau}{\bar{R}_vT_i^2}c_v,
\end{equation}
where for $\tau x\ll T_0$ we can replace $T_i$ with the {equilibrium} temperature $T_0$. In general, the solution to this equation yields an exponential concentration profile for $c_v$ and hence a nonlinear interfacial temperature profile \citep{Qin2015}. However, when the deviation of $c_v$ from $c_v^0$ is small, the concentration profile at the interface becomes approximately linear
\begin{equation} \label{eq:cvi}
\tilde{c}_v\approx\Omega \tilde{x},
\end{equation}
where
\begin{equation} \label{eq:Omega}
\Omega=c_v^0\frac{\mathcal{L}d_l\tau}{\bar{R}_vT_0^2}=c_v^0 \frac{H}{V} Ma,
\end{equation}
and
\begin{align}
\label{eq:H}
H=\frac{\mathcal{L}^2Dp_g^0}{\bar{R}_v^2T_0^3k_l}
\end{align}
is another nondimensional parameter, the meaning of which will become clear later. Again, since the product $Dp_g^0$ is independent of $c_a^0$, so is $H$.

\begin{figure}
\subfigure[]{\includegraphics[width=.32\columnwidth]{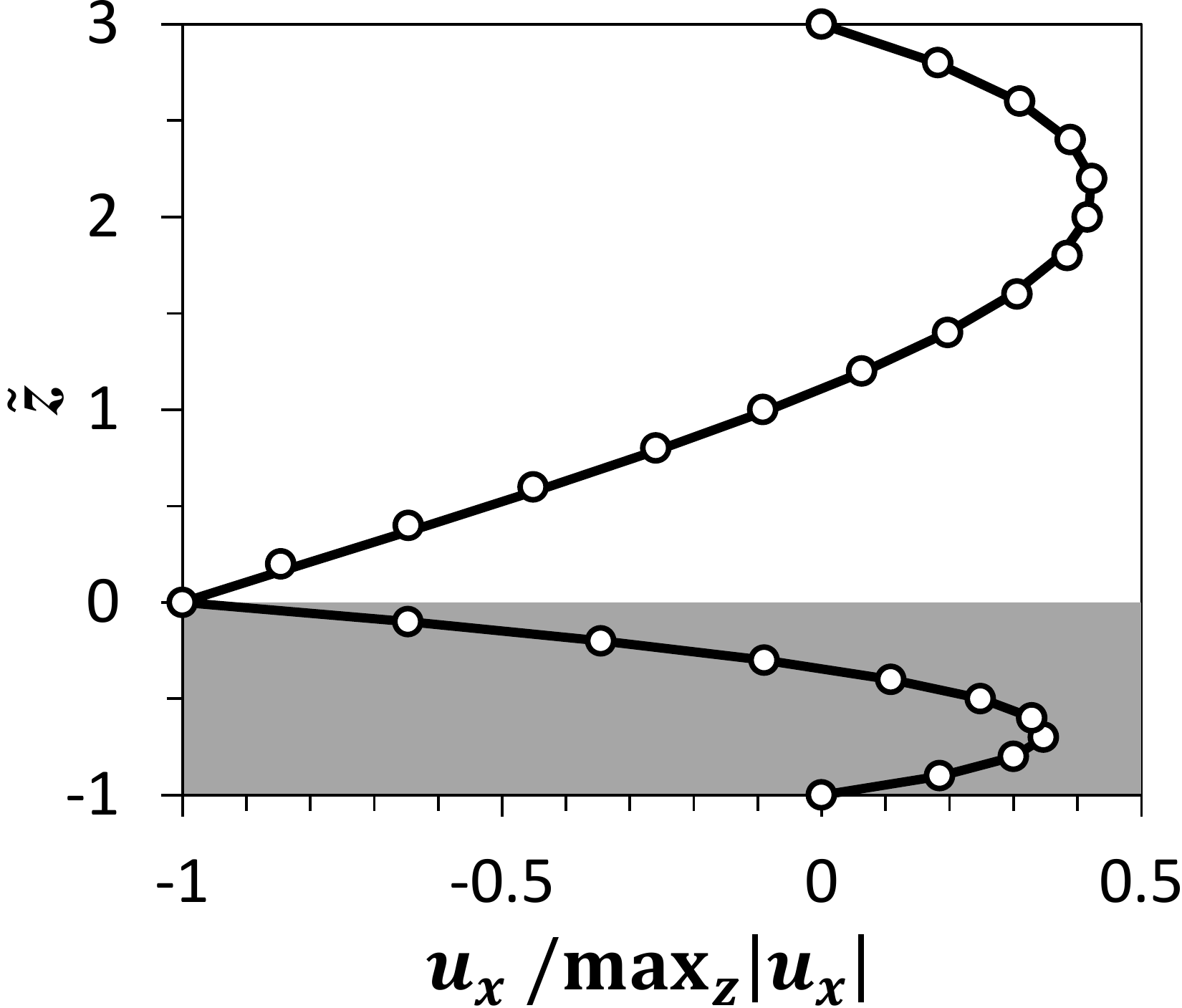}}\hspace{1mm}
\subfigure[]{\includegraphics[width=.32\columnwidth]{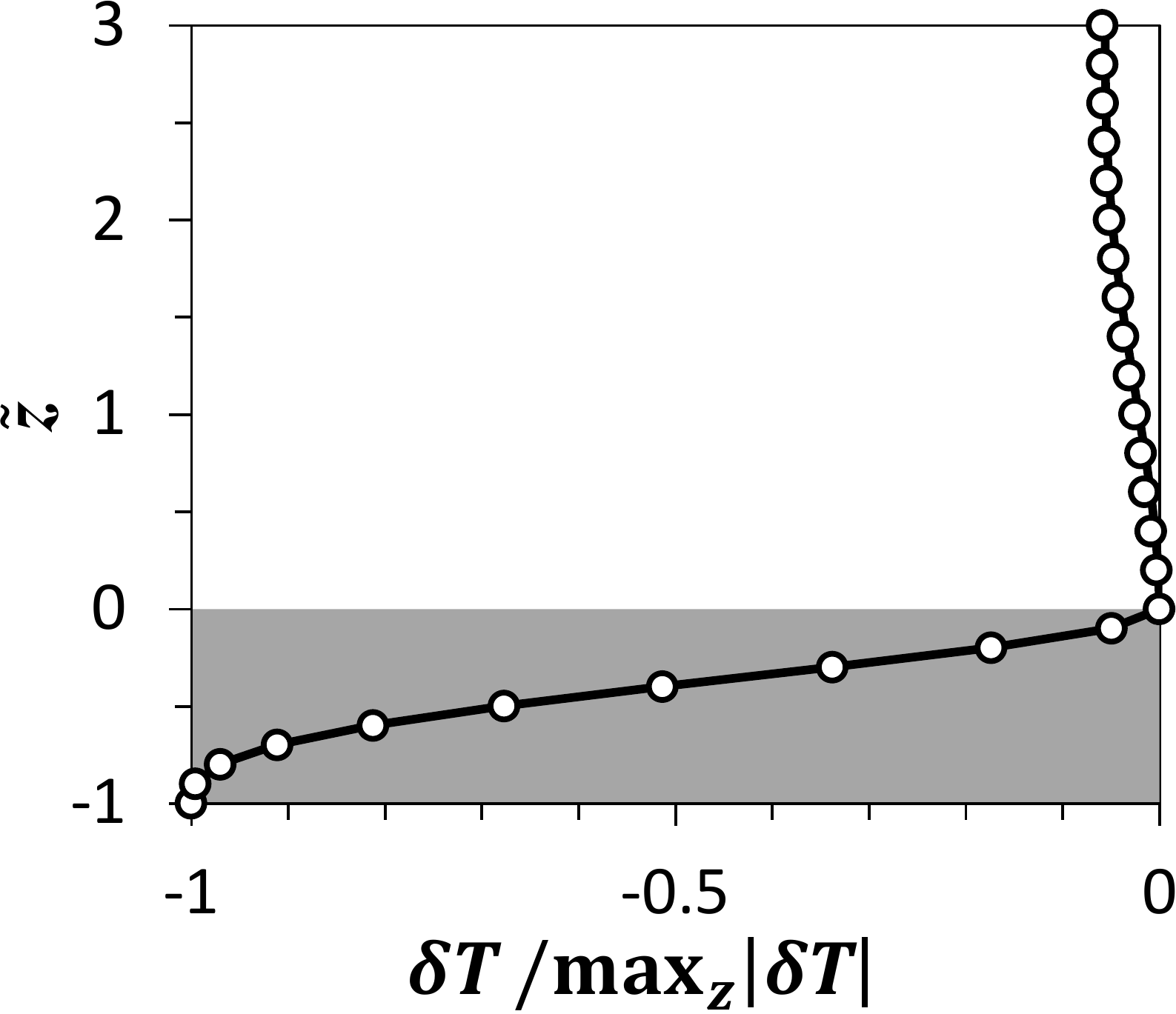}}\hspace{1mm}
\subfigure[]{\includegraphics[width=.32\columnwidth]{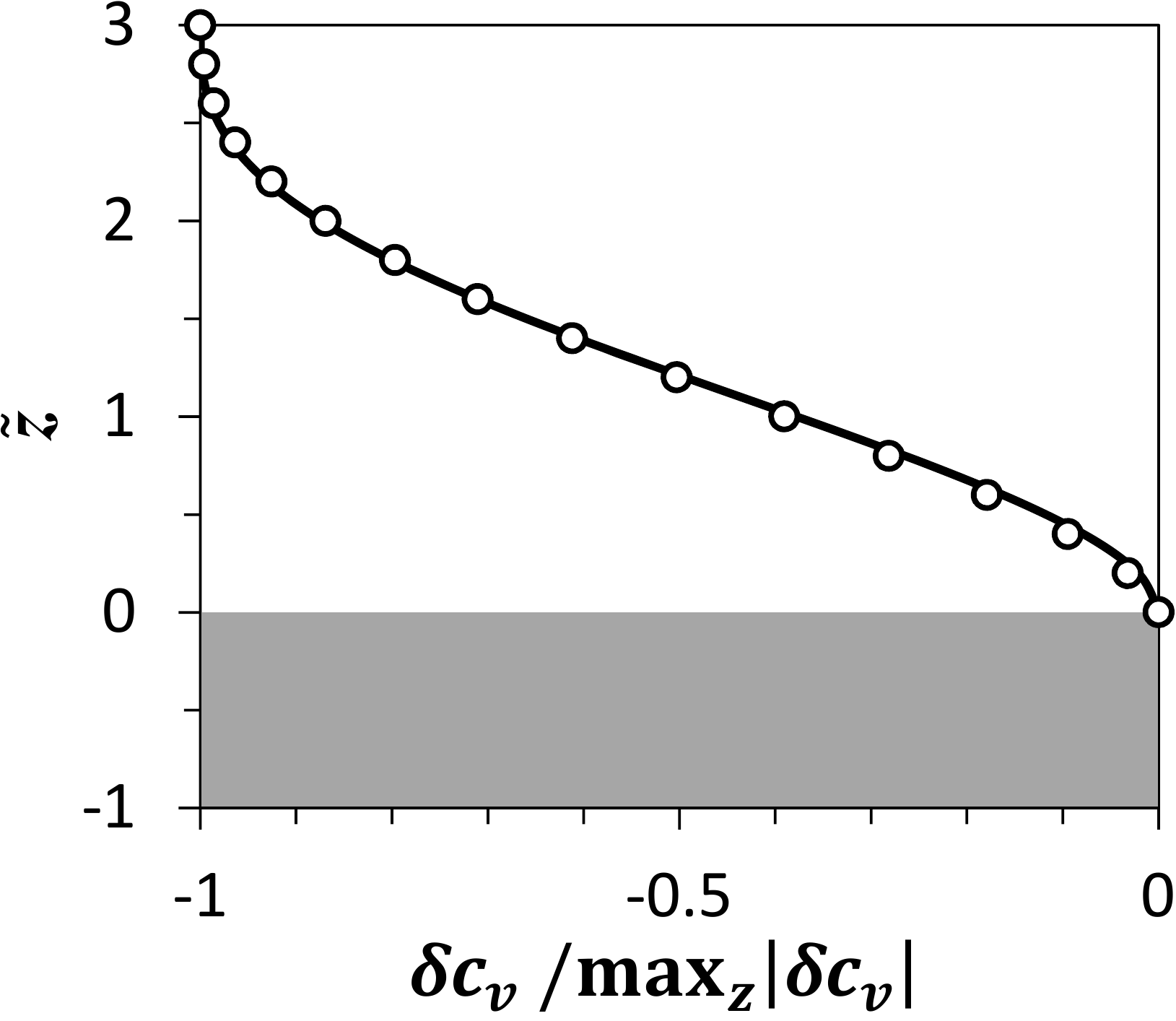}}
\caption{{Comparison between the numerical and analytical solutions for $Ma=190$ and $c_a^0 = 0.96$ (atmospheric conditions) in the middle of the cell, $x=L_x/2$. 
Shown are the normalized vertical profiles of (a) the horizontal velocity $u_{x} = u_x$, (b) the variation $\delta T=T-T_i$ of the temperature, and (c) the variation $\delta c_v=c_v-c_{v,i}$ of vapour concentration. Open circles correspond to numerical results and solid lines correspond to the analytical solutions.}
}
\label{fig:an_vs_num}
\end{figure}

Given the boundary conditions for $\tilde{T}_g$ and $\tilde{c}_v$, we can look for solutions to these two fields in the gas layer of the form
\begin{align}
\tilde{T}_g &= \tilde{x}+\tilde{\theta}_g(\tilde{z}),\nonumber\\
\tilde{c}_v &= \Omega[\tilde{x}+\tilde{\varsigma}_v(\tilde{z})],
\end{align}
where 
\begin{align}\label{eq:bcbot}
\tilde{\theta}_g(0)&=0,\nonumber\\
\tilde{\varsigma}_v(0)&=0.
\end{align}
For a uniform flow, $\tilde{\psi}_g$ should only depends on $\tilde{z}$, so the system (\ref{eq:ngas}) reduces to
\begin{align}\label{eq:baseeq3}
-\tilde{\psi}_g'''' + \Xi_T+\Xi_c\Omega &= 0,\nonumber\\
\tilde{\psi}_g' - K_\alpha\tilde{\theta}''_g&=0,\nonumber\\
\tilde{\psi}_g' - K_D\tilde{\varsigma}''_v&=0.
\end{align}
Solving these equations subject to the boundary conditions \eqref{eq:bctop}, \eqref{eq:bcvgb}, \eqref{eq:bcvgt}, and \eqref{eq:bcbot} at the interface and the top {of the gas layer} yields the steady state solutions for  the stream function
\begin{equation}
\label{eq:psi-gas}
\tilde{\psi}_g = -\mathcal{R}\left[\frac{\tilde{z}(\tilde{z}-A)^2}{4A^2} 
+ \mathcal{B}\frac{\tilde{z}(\tilde{z}-A)^2(2\tilde{z}+A)}{48A^3}\right],
\end{equation}
velocity
\begin{align}
\label{eq:utheo-gas}
\tilde{\bf u}_g = -\mathcal{R}\left[\frac{(\tilde{z}-A)(3\tilde{z}-A)}{4A^2}+\mathcal{B}\frac{(\tilde{z}-A)(8\tilde{z}^2-A\tilde{z}-A^2)}{48A^3}\right]\hat{x},
\end{align}
temperature
\begin{equation}
\label{eq:Tg-theo}
\tilde{T}_g=\tilde{x}-\frac{\mathcal{R}}{K_\alpha}\left[\frac{\tilde{z}^2(3\tilde{z}^2-8A\tilde{z}+6A^2)}{48A^2}+\mathcal{B}\frac{\tilde{z}^2(8\tilde{z}^3-15A\tilde{z}^2+10A^3)}{960A^3}\right],
\end{equation}
and vapour concentration in the gas phase
\begin{equation}
\label{eq:cv-theo}
\tilde{c}_v=\Omega\tilde{x}-\Omega\frac{\mathcal{R}}{K_D}\left[\frac{\tilde{z}^2(3\tilde{z}^2-8A\tilde{z}+6A^2)}{48A^2}+\mathcal{B}\frac{\tilde{z}^2(8\tilde{z}^3-15A\tilde{z}^2+10A^3)}{960A^3}\right].
\end{equation}
The parameters $\mathcal{R}$ and $\mathcal{B}$ are analogous to the Reynolds number and the dynamic Bond number, but incorporate the properties of both fluid layers:
\begin{align}
\label{eq:RB}
\mathcal{R}&=Re\left(1+\frac{Bo_D}{12}\right) + \frac{A^3}{12K_\nu}(\Xi_T+\Xi_\varsigma),
\nonumber\\
\mathcal{B}&=-\frac{A^3}{\mathcal{R}K_\nu}(\Xi_T+\Xi_\varsigma),
\end{align}
where we defined
\begin{align}\label{eq:Xi2}
\Xi_\varsigma=\Xi_c\Omega=
-\frac{(1-c_a^0) (\bar{R}_a-\bar{R}_v)}
 {\bar{R}_a-(\bar{R}_a-\bar{R}_v)c_a^0}
 {\frac{\mathcal{L}Bo_D}{\beta_l\bar{R}_v T^2}}\frac{Ma}{Pr}.
\end{align}
Note that the form of the analytical solutions (\ref{eq:utheo})-(\ref{eq:Ttheo}) and (\ref{eq:utheo-gas})-(\ref{eq:cv-theo}) is different from that of the solutions derived by \citet{Qin2013} because (i) the buoyancy force caused by the variation in the composition of the gas is explicitly taken into account in the present analysis, (ii) the mass transport in the gas phase is described using the {concentration rather than the mass density of the vapour}, and (iii) a different choice of the origin of the coordinate system is made. {Figure \ref{fig:an_vs_num} compares the analytical solutions with the numerical ones (cf. Fig. \ref{fig:SUF}) computed by \citet{Qin2015}. 
We find that the analytical solutions accurately describe the core region of the flow away from the end walls for sufficiently low $\tau$ at which the base flow is stable. }

\section{Linear Stability Analysis}
\label{sec:smf}

{The main focus of our study is on relatively thick layers where the effect of buoyancy is nonnegligible, such that $Bo_D=O(1)$.
In this parameter regime gravity is strong enough to keep the liquid interface nearly flat, so in our analysis we will assume that the thickness of the liquid layer remains uniform even after the onset of convection pattern.}
To date, only one study \citep{Priede1997} of pattern formation in buoyancy-thermocapillary convection correctly predicted the formation of a stationary pattern of convection rolls observed in experiment at $Bo_D=O(1)$. 
This study was based on a one-sided model which completely ignored phase change and assumed adiabatic conditions at the free surface. 
While such description may be acceptable for nonvolatile liquids or at high concentrations of air, it fails to describe volatile liquids at lower concentrations of air. 
Below a generalized analysis is presented, which accounts for both heat and mass flux across the interface associated with phase change and which is applicable regardless of the amount of air present in the {gas layer (quantified by $c_a^0$).} 

\subsection{Diffusion-Dominated Case}
\label{sec:a2}

While the nondimensional equations (\ref{eq:geq3}) for the liquid layer can be used without modification, in order to obtain a simplified problem, which can generate useful physical insight, we will start by making several approximations in the treatment of heat and mass transport on the gas side.
The relative contribution of advection and diffusion to the mass and heat transport in the gas layer are described by the P\'eclet numbers $Pe_m$ and $Pe_t$, respectively. 
Since the relevant length scale here is the wavelength of the convective pattern, which is comparable to the depth of the liquid layer, and the velocity scale is determined by the interfacial velocity $u_i=\nu_l\tilde{u}_l(0)/d_l$ (cf. Eq. \eqref{eq:utheo}), we have 
\begin{align}\label{eq:Pem}
Pe_m=\frac{|u_i|d_l}{D}=\frac{12+Bo_D}{48}\frac{Re}{K_D}
\end{align}
and
\begin{align}\label{eq:Pet}
Pe_t=\frac{|u_i|d_l}{\alpha_g}=\frac{12+Bo_D}{48}\frac{Re}{K_\alpha}.
\end{align}
In typical experiments \citep{Villers1992,riley1998,Li2013} both P\'eclet numbers are at most $O(1)$, so we can drop the advective terms {in the transport equations (\ref{eq:adv-cv}) and (\ref{eq:adv-T}) in the gas phase.
Furthermore, if we are only interested in the transition thresholds for stationary instability, we can also ignore the time derivatives, such that the system \eqref{eq:ngas} simplifies to}
\begin{align}
\tilde{\psi}_g&=0,\nonumber\\
(\partial_{\tilde{x}}^2+\partial_{\tilde{z}}^2)\tilde{T}_g&=0,\nonumber\\
(\partial_{\tilde{x}}^2+\partial_{\tilde{z}}^2)\tilde{c}_v&=0.\label{eq:diffg}
\end{align}
{Note} that neglecting the transient dynamics of the gas phase changes neither the critical Marangoni number nor the critical wavelength {of a (stationary) pattern}.

The base solution $\tilde{\psi}_{l0}$, $\tilde{T}_{l0}$ describing the uniform return flow in the liquid layer is given by (\ref{eq:psitheo})-(\ref{eq:Ttheo}).
The base solution to \eqref{eq:diffg} describing the gas layer in the diffusion-dominate case can be obtained by setting $\mathcal{R}=0$ in (\ref{eq:Tg-theo})-(\ref{eq:cv-theo}) which is equivalent to setting $\tilde{\bf u}_g=0$ ($\tilde{\psi}_g=0$) and gives:
\begin{align}
\tilde{T}_{g0}&=\tilde{x},\nonumber\\
\tilde{c}_{v0}&=\Omega\tilde{x}.\label{eq:baseg}
\end{align}


Although the base state satisfies the adiabatic boundary condition {at the interface}, perturbations in the temperature will give rise to heat and mass flux across the interface and, consequently, through the gas phase. Perturbed solutions can be written in the form of Fourier integrals 
\begin{align}
\tilde{\psi}_m &= \tilde{\psi}_{m0} +\int_{-\infty}^{\infty}{ \tilde{\psi}_{mq}(\tilde{z}) { e^{iqx+\sigma_q\,t}} dq},\nonumber\\
\tilde{T}_m &= \tilde{T}_{m0}+\int_{-\infty}^{\infty}{ \tilde{\theta}_{mq}(\tilde{z}) { e^{iqx+\sigma_q\,t}} dq},\nonumber\\
\tilde{c}_v &= \tilde{c}_{v0}+\Omega\int_{-\infty}^{\infty}{ \tilde{\varsigma}_{vq}(\tilde{z}) { e^{iqx+\sigma_q t}} dq},
\end{align}
where $m=\{l,g\}$ denotes the phase, $q$ is the wavenumber describing the variation in the horizontal direction, and $\sigma_q$ is the temporal growth rate. The functions $\tilde{\psi}_{mq} (\tilde{z})$, $\tilde{\theta}_{mq}(\tilde{z})$, and $\tilde{\varsigma}_{vq}(\tilde{z})$ define the vertical profile for the perturbations in, respectively, the stream function $\tilde{\psi}$, temperature $\tilde{T}$, and vapour concentration $\tilde{c}_v$ with wavenumber $q$.

Rewriting (\ref{eq:diffg}) in terms of the perturbations, we obtain
\begin{align}
\tilde{\psi}_{gq} &=  0,\nonumber \\
\tilde{\theta}_{gq}'' &=  q^2 \tilde{\theta}_{gq},\nonumber \\
\tilde{\varsigma}_{vq}'' &=  q^2 \tilde{\varsigma}_{vq}. \label{eq:per1}
\end{align}
Temperature continuity at the interface requires
\begin{align}
\label{eq:bctheta}
\tilde{\theta}_{gq}(0) = \tilde{\theta}_{lq}(0).
\end{align}
At the top of the gas layer we have
\begin{align}
\tilde{\theta}_{gq}'(A) &= 0,\nonumber \\
\tilde{\varsigma}_{vq}'(A) &= 0. \label{eq:bctheta2}
\end{align}
The solution of \eqref{eq:per1} satisfying these boundary conditions is
\begin{align}\label{eq:thetagz}
\tilde{\theta}_{gq}(\tilde{z}) = \frac{\cosh\left[q (\tilde{z} - A)\right]}{\cosh\left( q A \right)}\tilde{\theta}_{gq}(0).
\end{align}
The solution for the perturbation in the vapour concentration is analogous,
\begin{align}
\label{eq:cvq}
\tilde{\varsigma}_{vq}(\tilde{z}) = \frac{\cosh\left[q(\tilde{z} - A)\right]}{\cosh\left(q A \right)}\tilde{\varsigma}_{vq}(0),
\end{align}
where $\tilde{\varsigma}_{vq}(0)$ can be related to the perturbation in the interfacial temperature via the Clausius-Clapeyron relation \eqref{eq:CC}
\begin{align}\label{eq:rhoviq}
\tilde{\varsigma}_{vq}(0) = \tilde{\theta}_{gq}(0).
\end{align}

Fourier transforming (\ref{eq:jcv}) yields
\begin{align}
\label{eq:j1q}
\tilde{J}_q= -\frac{\Omega}{c_a^0}\tilde{\varsigma}'_{vq}(0),
\end{align}
where $\tilde{J}_q$ is the Fourier coefficient of $\tilde{J}(\tilde{x})$. Subtracting the base solutions from the heat balance (\ref{eq:bc4}), Fourier transforming the result, and using (\ref{eq:j1q}) gives the following relation
\begin{align}
\label{eq:qper}
\tilde{\theta}'_{lq}(0)&= \frac{k_g}{k_l} \tilde{\theta}'_{gq}(0) + \frac{\Omega}{c_a^0}\frac{V}{Ma}\tilde{\varsigma}'_{vq}(0).
\end{align}
With the help of (\ref{eq:thetagz}), (\ref{eq:cvq}), and (\ref{eq:rhoviq}) this can rewritten in the form of Newton's cooling law
\begin{align}
\label{eq:qper2}
\tilde{\theta}'_{lq}(0) &=-Bi_q \tilde{\theta}_{lq}(0),
\end{align}
where we introduced a wavenumber-dependent analogue of the Biot number
\begin{equation}\label{eq:Bi2}
Bi_q  = q\tanh\left(q A \right)\left[\frac{k_g}{k_l} + \frac{1-c_a^0}{c_a^0}H \right],
\end{equation}
which will be referred to as the Biot coefficient below to highlight the fact that it is a function of $q$ and $c_a^0$, not a constant. 
The prefactor $\tanh(q A)$ describes the effect of finite thickness of the gas layer. For $d_g\gg d_l$ ($A\gg 1$), it approaches unity and (\ref{eq:Bi2}) reduces to the expression derived by Chauvet {\it et al.} \citep{Chauvet2012} in the context of stability of a volatile liquid layer in the presence of a {\it vertical} temperature gradient. The first term describes the effect of thermal conduction through the gas layer, while the second term describes the effect of latent heat released/absorbed at the interface as a result of condensation/evaporation.


By linearizing the governing equations (\ref{eq:nliq}) around the base state \eqref{eq:psitheo}-\eqref{eq:Ttheo}, we obtain the evolution equations for the perturbations $\tilde{\psi}_{lq}$ and $\tilde{\theta}_{lq}$ in the liquid layer
\begin{align}
\tilde{\nabla}_q^2\tilde{\nabla}_q^2\tilde{\psi}_{lq} + iq C_1(\tilde{z})Re \tilde{\psi}''_{lq}  - iq C_2(\tilde{z}) Re \tilde{\psi}_{lq} - iq Gr  \tilde{\theta}_{lq} &  =\sigma_q\tilde{\nabla}_q^2\tilde{\psi}_{lq},\nonumber\\
\label{eq:perl}
Pr^{-1}\tilde{\nabla}_q^2\tilde{\theta}_{lq} + iqC_1(\tilde{z})Re\tilde{\theta}_{lq} - iqC_3(\tilde{z})Ma \tilde{\psi}_{lq} - \tilde{\psi}'_{lq}&= \sigma_q\tilde{\theta}_{lq},
\end{align}
where we defined $\tilde{\nabla}_q^2=\partial_{\tilde{z}}^2-q^2$ and
\begin{align}
C_1(\tilde{z}) &=\frac{(\tilde{z}+1)(3\tilde{z}+1)}{4} - Bo_D\frac{(\tilde{z}+1)(8\tilde{z}^2+\tilde{z}-1)}{48},\nonumber\\
C_2(\tilde{z}) &=q^2C_1(\tilde{z})+\frac{3}{2}-Bo_D\frac{8\tilde{z}+3}{8},\nonumber\\
C_3(\tilde{z}) &=\frac{\tilde{z}(\tilde{z}+1)^2}{4} - Bo_D\frac{\tilde{z}(\tilde{z}+1)^2(2\tilde{z}-1)}{48}.
\end{align}
This is a system of ODEs which is fourth-order with respect to $\tilde{\psi}_{lq}$ and second-order with respect to $\tilde{\theta}_{lq}$ and hence needs a total of six boundary conditions. These  boundary conditions are:
\begin{align}
\tilde{\psi}_{lq}(-1)&=0,\nonumber \\
\tilde{\psi}'_{lq}(-1)&=0,\nonumber \\
\tilde{\theta}'_{lq}(-1) &= 0.\label{eq:bcn1}
\end{align}
at the bottom of the liquid layer and
\begin{align}
\tilde{\psi}_{lq}(0)&=0,\nonumber \\
\tilde{\psi}''_{lq}(0) &= -i q Re \tilde{\theta}_{lq}(0)\label{eq:bcn6}
\end{align}
and
\begin{align}
\tilde{\theta}'_{lq}(0) &= -Bi_q \tilde{\theta}_{lq}(0) \label{eq:bcn7}
\end{align}
at the free surface. In the subsequent discussion we will refer to the system \eqref{eq:perl} with the boundary conditions \eqref{eq:bcn1}--\eqref{eq:bcn7} as the enhanced one-sided model, since it incorporates the effect of the gas phase entirely through boundary conditions at the interface.

\subsection{Transient Dynamics in the Gas Layer}
\label{sec:a5}

A more accurate description of the instability, {stationary or oscillatory,} can be obtained by restoring the time dependence of the temperature and composition of the gas phase. This corresponds to replacing the Laplace equations (\ref{eq:diffg}) with 
\begin{align}
\partial_{\tilde{t}} \tilde{T}_g&=K_\alpha\tilde{\nabla}^2\tilde{T}_g,\nonumber\\
\label{eq:geqg4}
\partial_{\tilde{t}} \tilde{c}_v&=K_D\tilde{\nabla }^2\tilde{c}_v.
\end{align}
The corresponding equations for the perturbations are
\begin{align}
\tilde{\psi}_{gq}&=0,\nonumber\\
\tilde{\theta}_{gq}''-q^2 \tilde{\theta}_{gq}&=\sigma_q K_\alpha^{-1}\tilde{\theta}_{gq},\nonumber\\
\label{eq:per2}
\tilde{\varsigma}_{vq}''-q^2 \tilde{\varsigma}_{vq}&=\sigma_q K_D^{-1}\tilde{\varsigma}_{vq}.
\end{align}
These equations should be solved subject to the boundary conditions \eqref{eq:bctheta}, \eqref{eq:bctheta2}, \eqref{eq:rhoviq}, \eqref{eq:bcn1}, \eqref{eq:bcn6} and
\begin{align}
\label{eq:qper3}
\tilde{\theta}_{lq}'(0)&= -\frac{k_g}{k_l}\bar{\theta}'_{gq}(0)
- \frac{1-c_a^0}{c_a^0}H \bar{\varsigma}'_{vq}(0),
\end{align}
which follows from the heat flux balance (\ref{eq:bc4}) and the mass flux balance (\ref{eq:j1q}) and
replaces \eqref{eq:qper2}.

{Since it involves an incomplete description of the gas layer (the effects of advection are not included), this model based on the governing equations \eqref{eq:perl} and \eqref{eq:per2} can be called ``one-and-a-half-sided.''}
Note that, since both $K_\alpha$ and $K_D$ are typically large compared with unity, the terms on the right-hand-side of (\ref{eq:per2}) are small. Dropping them would lead to the {enhanced one-sided model} derived in Section \ref{sec:a2}. 

\subsection{The Effect of Advection in the Gas Layer}
\label{sec:a6}

The effect of advection can also be incorporated for the range of $c_a^0$ where the base solution \eqref{eq:psi-gas}-\eqref{eq:cv-theo} is valid. By setting $\tilde{\psi}_{gq}\ne 0$ we ensure that all of the transport mechanisms in the gas phase are accounted for and both layers are treated using an equally comprehensive model. The linearized evolution equations for the perturbations $\tilde{\psi}_{gq}$, $\tilde{\varsigma}_{vq}$, and $\tilde{\theta}_{gq}$ in the gas phase are
\begin{align}
{K_\nu}\tilde{\nabla}_q^2\tilde{\nabla}_q^2\tilde{\psi}_{gq} + iq \tilde{C}_1(\tilde{z})\mathcal{R} \tilde{\psi}''_{gq}  - iq \tilde{C}_2(\tilde{z}) \mathcal{R} \tilde{\psi}_{gq} - iq \Xi_T \tilde{\theta}_{gq} - iq\Xi_\varsigma \tilde{\varsigma}_{vq} &  =\sigma_q\tilde{\nabla}_q^2\tilde{\psi}_{gq},\nonumber\\
K_\alpha\tilde{\nabla}_q^2\tilde{\theta}_{gq} + iq\tilde{C}_1(\tilde{z})\mathcal{R}\tilde{\theta}_{gq} - iq\tilde{C}_3(\tilde{z}) \mathcal{R} K_\alpha^{-1}\tilde{\psi}_{gq} - \tilde{\psi}'_{gq}&= \sigma_q\tilde{\theta}_{gq},
\nonumber\\
\label{eq:per3}
K_D\tilde{\nabla}_q^2\tilde{\varsigma}_{vq} + iq\tilde{C}_1(\tilde{z})\mathcal{R}\tilde{\varsigma}_{vq} - iq\tilde{C}_3(\tilde{z}) \mathcal{R} K_D^{-1}\tilde{\psi}_{gq} - \tilde{\psi}'_{gq}&= \sigma_q\tilde{\varsigma}_{vq}.
\end{align}
where
\begin{align}
\tilde{C}_1(\tilde{z}) &=\frac{(\tilde{z}-A)(3\tilde{z}-A)}{4A^2} 
+ \mathcal{B}\frac{(\tilde{z}-A)(8\tilde{z}^2-A\tilde{z}-A^2)}{48A^3},\nonumber\\
\tilde{C}_2(\tilde{z}) &=q^2\tilde{C}_1(\tilde{z}) +\frac{3}{2A^2}
+ \mathcal{B}\frac{8\tilde{z}-3A}{8A^3},\nonumber\\
\tilde{C}_3(\tilde{z}) &=\frac{\tilde{z}(\tilde{z}-A)^2}{4A^2} 
+ \mathcal{B}\frac{\tilde{z}(\tilde{z}-A)^2(2\tilde{z}+A)}{48A^3}.
\end{align}

The boundary conditions are given by \eqref{eq:bctheta}, \eqref{eq:bctheta2}, \eqref{eq:rhoviq}, \eqref{eq:bcn1}, \eqref{eq:bcn6}, and \eqref{eq:qper3}. In addition, the boundary conditions for $\bar{\psi}_{gq}(\bar{z})$ follow from (\ref{eq:bcvgb}) and (\ref{eq:bcvgt}):
\begin{align}
\tilde{\psi}_{gq}(A)&=0,\nonumber \\
\tilde{\psi}'_{gq}(A)&=0
\end{align}
at the top of the gas layer and
\begin{align}
\tilde{\psi}_{gq}(0)&=0,\nonumber \\
\tilde{\psi}'_{gq}(0) &= \tilde{\psi}'_{lq}(0)
\end{align}
at the free surface. In the subsequent discussion we will refer to the system \eqref{eq:perl} and \eqref{eq:per3} with the appropriate boundary conditions as the two-sided model, since it treats both the liquid and the gas phase with the same level of detail.

\subsection{Different Modes of Instability}

{
The boundary value problem, whether it is described by the system \eqref{eq:perl} by itself or combined with \eqref{eq:per2} or \eqref{eq:per3}, has a spectrum of eigenvalues $\sigma_q^n$ and the corresponding eigenfunctions $\{\psi_{lq}^n,\theta_{lq}^n,\cdots\}$, $n=0,1,\cdots$, for a given complex wavenumber $q=k+is$ and Marangoni number $Ma$.
The stability of the base flow is determined by the eigenvalue $\sigma_q^0=\kappa_q+i\omega_q$ with the largest real part $\kappa_q$ and the character of the instability (oscillatory or stationary) is determined by the imaginary part $\omega_q$. 
}

{
So far we have only focused on the boundary conditions in the vertical (confined) direction.
In both experiment and simulations the fluid layers are also bounded in the horizontal (extended) directions $0<x<L_x$ and $0<y<L_y$.
Comparison of various experimental results suggests that the spanwise aspect ratio $\Gamma_y=L_y/d_l$ is not particularly important: aside from the orientation of hydrothermal waves at low $Bo_D$, the pattern that emerges above the threshold of primary instability is basically the same for both $\Gamma_y=O(1)$ \citep{Villers1992,Desaedeleer1996,Garcimartin1997,Li2013} and $\Gamma_y\gg 1$ \citep{riley1998,Burguete2001}.
This is not very surprising since the boundaries in the spanwise direction play a passive role --  they tend to suppress, rather than enhance convection.
}

{
The boundaries in the streamwise direction are less trivial and play a crucial role in selecting the convection pattern for any $\Gamma_x=L_x/d_l$.
The corresponding boundary conditions are of the enhancing type \citep{Cross2009}: buoyancy, which plays a significant role for $Bo_D=O(1)$, drives an upward flow near the hot end wall, generating a single localized convection roll.
Similarly, buoyancy drives a downward flow near the cold end wall, generating another, much weaker, convection roll. 
These rolls appear for $Ma$ well below the critical value $Ma_c$ at which convection pattern emerges away from the end walls.
The rolls are steady over a wide range of $Ma$ and have a horizontal extent of $O(d_l)$.
For $\Gamma_x\gg1$ and sufficiently low $Ma$, the flow in the core region between these two convection rolls, i.e., for $O(d_l)<x<L_x-O(d_l)$, is described well by the base flow solution \eqref{eq:utheo} and \eqref{eq:utheo-gas}, as mentioned previously (cf. Fig. \ref{fig:SUF}).
\citet{riley1998} refer to this state as a steady unicellular flow (SUF).
Since the perturbation strength is always finite at the end walls, defining the threshold of the primary instability at finite $\Gamma_x$ requires some care. 
}

\subsubsection{Convective Instability}

{
The dominant mode of primary instability in laterally unbounded systems (infinite $\Gamma_x$) tends to be convective.
The threshold of convective instability corresponds to $\kappa_q$ being negative over all $q$-real, except for $q=k_c$, where $k_c$ is the (real) critical wavenumber which defines the wavelength $\lambda/d_l=2\pi/k_c$ of the pattern that develops above threshold.
Therefore, the critical wavenumber $k_c$ and the critical Marangoni number $Ma_c$ are given by the  solution of the two equations
\begin{align}
\kappa_q|_{q_c}&=0,\nonumber\\
\left.\frac{\partial\kappa_q}{\partial q}\right|_{q_c}&=0, \label{eq:sigma_conv}
\end{align}
over $q$-real.
Generally, the critical frequency $\omega_c=\omega_q|_{q_c}\ne 0$, so the resulting pattern will be oscillatory, with disturbances amplified in the frame of reference moving with the group velocity \citep{Huerre2000} 
\begin{align}
v_g=-\left.\frac{\partial\omega_q}{\partial q}\right|_{q_c}. \label{eq:v_conv}
\end{align}
}

{
In laterally bounded systems, such as the fluid volumes with finite aspect ratios $\Gamma_x$ discussed in this paper, convective instabilities should not be observed in the absence of a localized oscillatory source of the right temporal frequency ($\omega\approx\omega_c$). 
An example of such a source could be an oscillating convection roll near the left (right) end wall for $v_g>0$ ($v_g<0)$ driven by a completely different mechanism (e.g., buoyancy due to cooling or heating at the end wall). 
While such oscillating convection rolls have indeed been observed in experiment \citep{Desaedeleer1996,Garcimartin1997,Li2013} and simulations \citep{Qin2013}, they arise {\it after} convection pattern in the core region of the flow has already been established.
Hence, this is a secondary, rather than a primary, instability. We should only expect to see a primary convective instability in systems with very large streamwise aspect ratio $\Gamma_x$ where infinitesimal disturbances have sufficient time $\Gamma_x/v_g$ to develop.
}

\subsubsection{Absolute Instability}

{
Absolute instabilities arise when infinitesimal disturbances grow in the laboratory frame.
Since they do not need a finite-amplitude source, they can arise both in laterally unbounded systems as well as in bounded systems with moderate $\Gamma_x$.
Absolute instability requires that the group velocity $v_g=0$ at onset, yielding a system of three real equations for three real unknown $k_c=\mathrm{Re}(q_c)$, $s_c=\mathrm{Im}(q_c)$, and $Ma_c$, which can be rewritten as a system of two equations \citep{Huerre2000}
\begin{align}
\kappa_q|_{q_c}&=0,\nonumber\\
\left.\frac{\partial\sigma_q^0}{\partial q}\right|_{q_c}&=0, \label{eq:sigma_abs}
\end{align}
the second of which is complex, for $q_c$-complex and $Ma_c$.
}

{
The resulting pattern will generally be oscillatory, $\omega_c\ne 0$, just like in the case of convective instability.
Although its group velocity vanishes, the phase velocity $v_p=-\omega_c/k_c$ of the pattern will be nonzero.
A good example of such an absolute instability is hydrothermal waves \citep{Smith1983a,Smith1983b} that are observed in the limit of low $Bo_D$ \citep{Daviaud1993,riley1998,Burguete2001}.
In the following, where necessary to avoid confusion, we will use the superscript $a$ to denote the critical values corresponding to absolute instability.
}

\subsubsection{Stationary Instability}

{
The threshold of a stationary instability corresponds to both the real and imaginary part of $\sigma_q^0$ vanishing 
\begin{align}
&\kappa_q|_{q_c}=0,\nonumber\\
&\omega_q|_{q_c}=0. \label{eq:sigma_stat}
\end{align}
These two equations should be solved for the critical wavenumber $k_c$ and the critical Marangoni number $Ma_c$, which both depend on the spatial attenuation rate $s$. 
In the following, where necessary to avoid confusion, we will use the superscript $s$ to denote the critical values corresponding to stationary instability.
In particular, we will write $Ma_c^s(s)$ to explicitly denote the dependence on $s$.
}

{
In laterally infinite systems the transition threshold is well-defined and corresponds to a spatially uniform stationary pattern ($s=0$).
Such a pattern will be referred to as the steady multicellular (SMC) state below. 
For $Ma<Ma_c^s(0)$ the convection pattern arises due to forcing at one of the boundaries (convection rolls driven by buoyancy, cf. Fig. \ref{fig:SUF}(c) and (d)), rather than spontaneously, and generally does not extend across the entire length of the system.
In this case the absolute value of $s=\mathrm{Im}(q_c)$ defines the spatial attenuation of the perturbation (or the number of convection rolls that can be detected in the liquid layer), while $k_c=\mathrm{Re}(q_c)$ determines the wavelength $\lambda$ of the pattern (or the distance between convection rolls).
For $s\ne 0$ the pattern appears below $Ma_c^s(0)$ and is localized near the right (hot) end wall for $s<0$ or near the left (cold) end wall for $s>0$.
As we will see below, $dMa_c^s/ds>0$, which means that the pattern always appears near the hot end wall (cf. Fig. \ref{fig:SUF}(d)) and we can only consider $s\le 0$.
}

{
When $e^{-|s|\Gamma_x}\ll 1$, the pattern extends over a region of length $O(2d_l/|s|)<L_x$. We will refer to this as partial multicellular (PMC) state following \citet{Qin2015} and \citet{Li2013}.
In this analysis, we set $s_\mathrm{PMC} = -1$ to define the transition from SUF to PMC at $Ma_c^s(-1)<Ma_c^s(0)$, which roughly corresponds to one extra roll nucleating near the (hot) end wall.
As $Ma$ is increased beyond this value, the convection pattern extends further away from the end wall, until it eventually covers the entire fluid layer. 
In practice this roughly corresponds to $e^{-|s|\Gamma_x}\approx 0.1$ or $s_\mathrm{SMC}= \pm2/\Gamma_x$. 
This is the definition that has to be used for $Pr\lesssim 1$ (i.e., for gases or metallic liquids featuring sensitive dependence of $Ma_c^s$ on $s$ for small $|s|$, as shown below). 
For typical (nonmetallic) liquids with $Pr=O(10)$, we can set $s_\mathrm{SMC}=0$, as in the case of laterally unbounded layers, as long as $\Gamma_x\gtrsim 20$. 
This is the case in both the experimental \citep{riley1998,Li2013} and numerical studies \citep{Qin2013,Qin2015} used for comparison below.
}

Note that, for $\sigma_q^0=0$, equations (\ref{eq:per2}) reduce to \eqref{eq:per1} even if $K_\alpha$ and $K_D$ are not large. Hence the two formulations always give identical predictions for the threshold of stationary instability (even though they predict different growth rates for modes with wavenumbers $q$ different from the critical one). Therefore, in what follows we will only focus on the limits when either both $Pe_m$ and $Pe_t$ are vanishingly small and stability is described by \eqref{eq:perl} alone, or when they are both $O(1)$, so \eqref{eq:perl} and \eqref{eq:per3} have to be solved simultaneously.

\section{Results}
\label{sec:res}

Previous theoretical studies of stability focused mainly on the dependence of the critical Marangoni number $Ma$ on the dynamic Bond number $Bo_D$ and Prandtl number $Pr$ characterizing the liquid layer. 
We have also investigated how both the critical $Ma$ and the critical wavelength $\lambda$ depend on the {equilibrium} concentration $c_a^0$ of air {in the gas layer}.
Where possible, the predictions of linear stability analysis have been compared with the available experimental and numerical data.
{
The boundary value problems, which determine $\sigma_q^0$ as a function of the wavenumber $q$ and various nondimensional parameters, were solved using the function {\bf bvp5c} in Matlab 2016a (cf. Appendix \ref{sec:numbvp} for details) and the equations \eqref{eq:sigma_conv}, \eqref{eq:sigma_abs}, and \eqref{eq:sigma_stat}, defining the conditions for various types of instabilities where solved using the Newton method.
}

\subsection{Convection at atmospheric conditions}

{
The most comprehensive experimental study to date that investigated convection patterns in the problem considered here is due to \citet{riley1998}, who determined the dependence of both the critical $Ma$ and the critical wavelength $\lambda$ corresponding to the onset of various instabilities on the dynamic Bond number $Bo_D$ for the 1 cSt silicone oil with  $Pr=13.9$. 
Since the authors have not reported the values of all material parameters of the liquid, we took the missing ones from \citet{Kavehpour2002}. 
No data has been found for the material properties of the vapour; these were assumed to be the same as those of the 0.65 cSt silicone oil \citep{yaws2003,yaws2009}. 
Any potential uncertainty in the values of these material parameters, however, will have a negligible effect on the results.
The 1.0 cSt silicone oil is not particularly volatile (the saturation vapour pressure is $p_v = 0.5$ kPa at $T = 298$ K, which corresponds to $c_a^0 = 0.995$ at atmospheric conditions), therefore the properties of the gas phase are essentially those of air.
}

\begin{figure}
\subfigure[]{\includegraphics*[width=0.49\columnwidth]{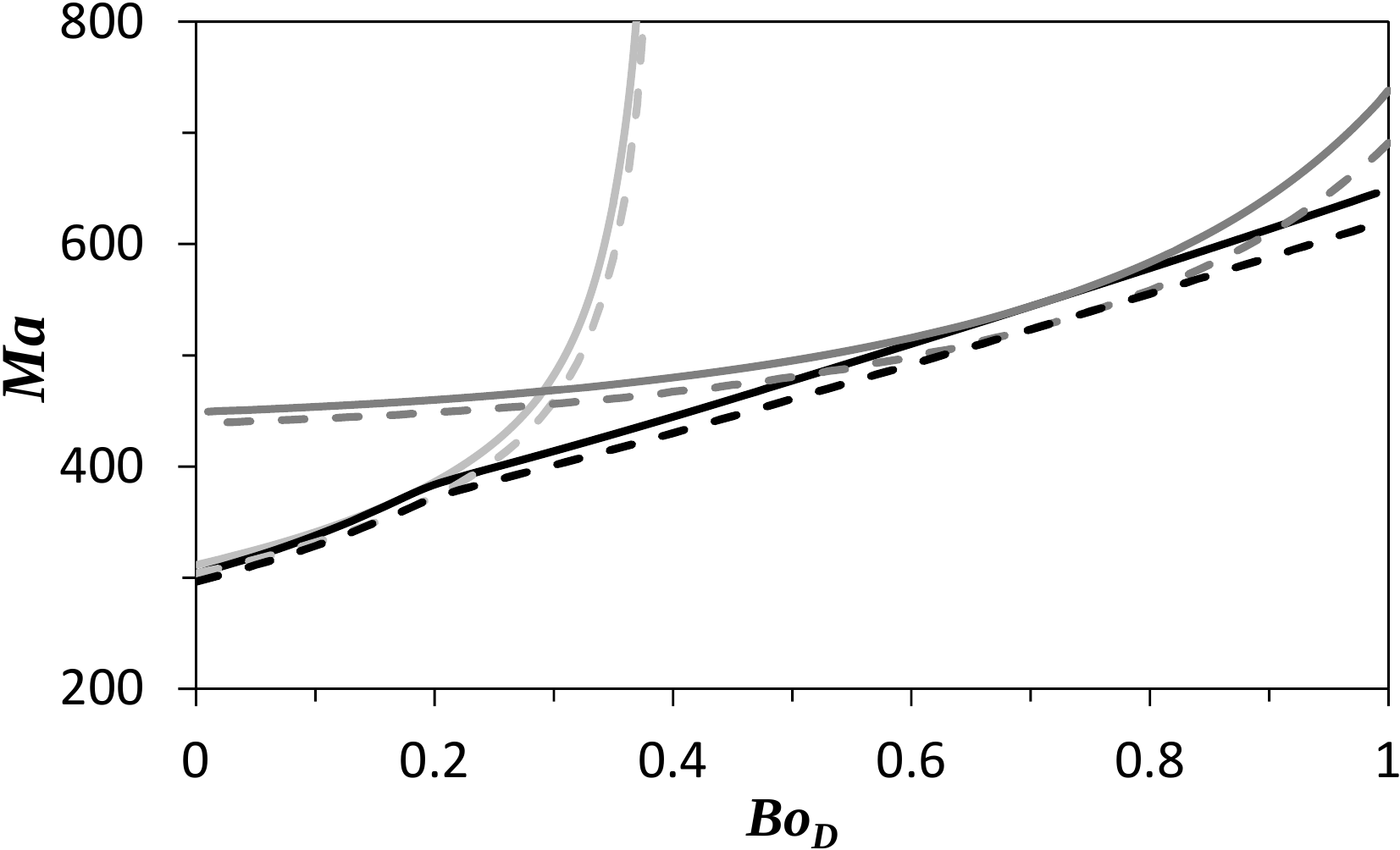}}\hspace{1mm}
\subfigure[]{\includegraphics*[width=0.49\columnwidth]{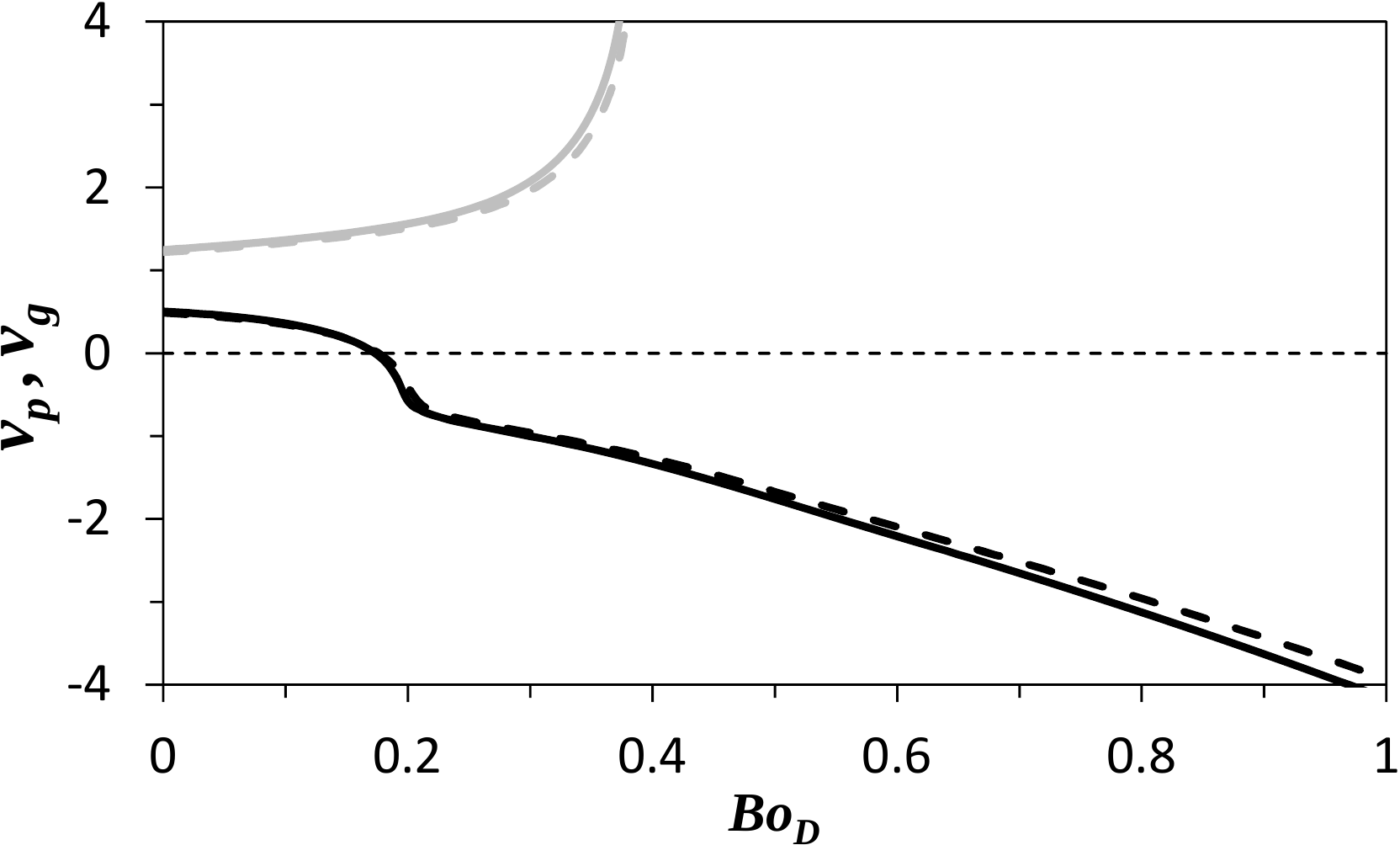}}
\caption{{Different types of instability in a laterally infinite layer of 1.0 cSt silicone oil at atmospheric conditions ($c_a^0=0.995$) as a function of the dynamic Bond number.
(a) Critical $Ma$ for the absolute (light gray), stationary (dark gray), and convective (black) instability. 
(b) Group velocity $v_g$ for convective instability (black) and phase velocity $v_p$ for absolute instability (light gray). 
Analytical predictions of the two-sided model are shown as solid lines and the predictions of the one-sided model with $Bi_q=0$ as dashed lines.}
}
\label{fig:Ma-RN-inf}
\end{figure}

\begin{figure}
\subfigure[]{\includegraphics*[width=0.49\columnwidth]{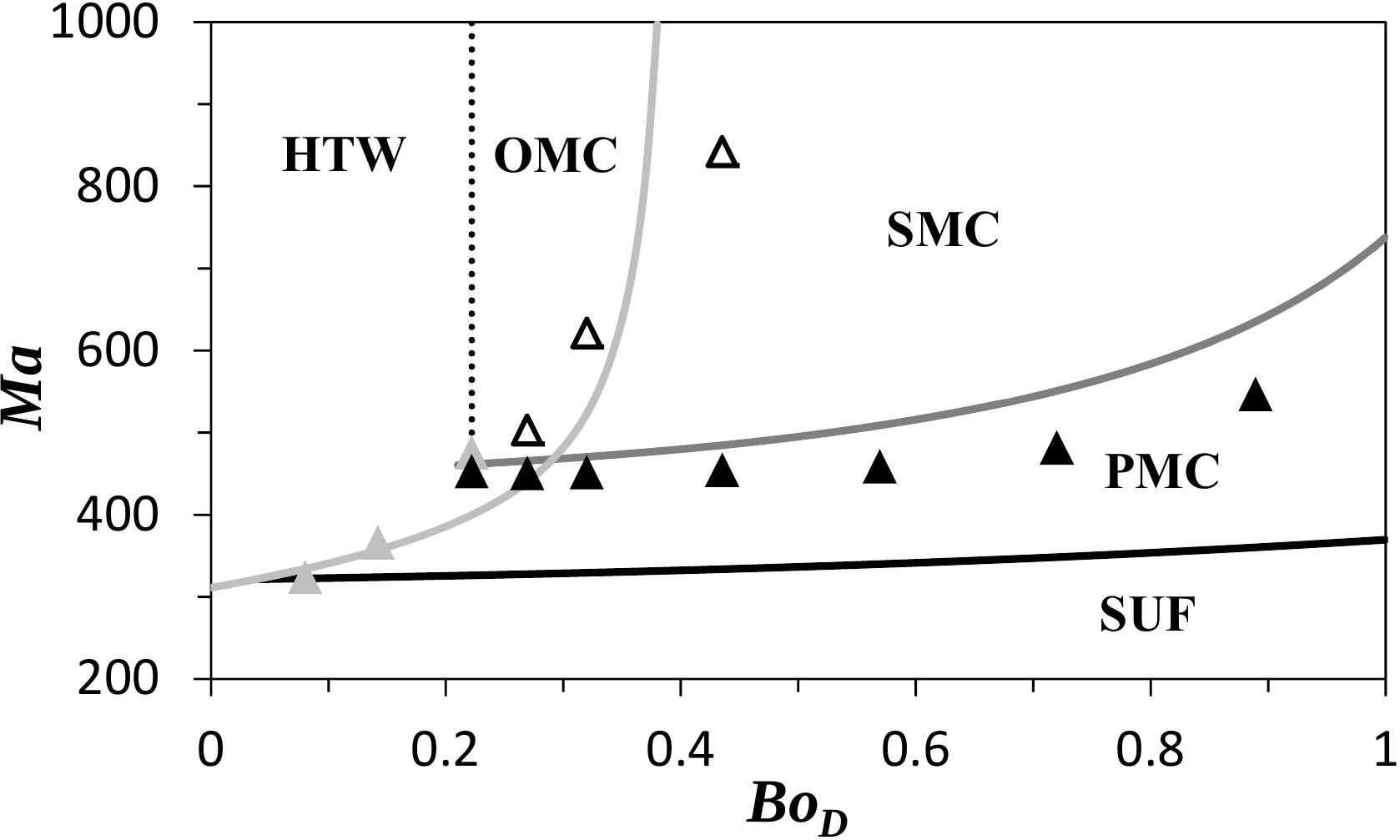}\vspace{-1mm}}\hspace{1mm}
\subfigure[]{\includegraphics*[width=0.49\columnwidth]{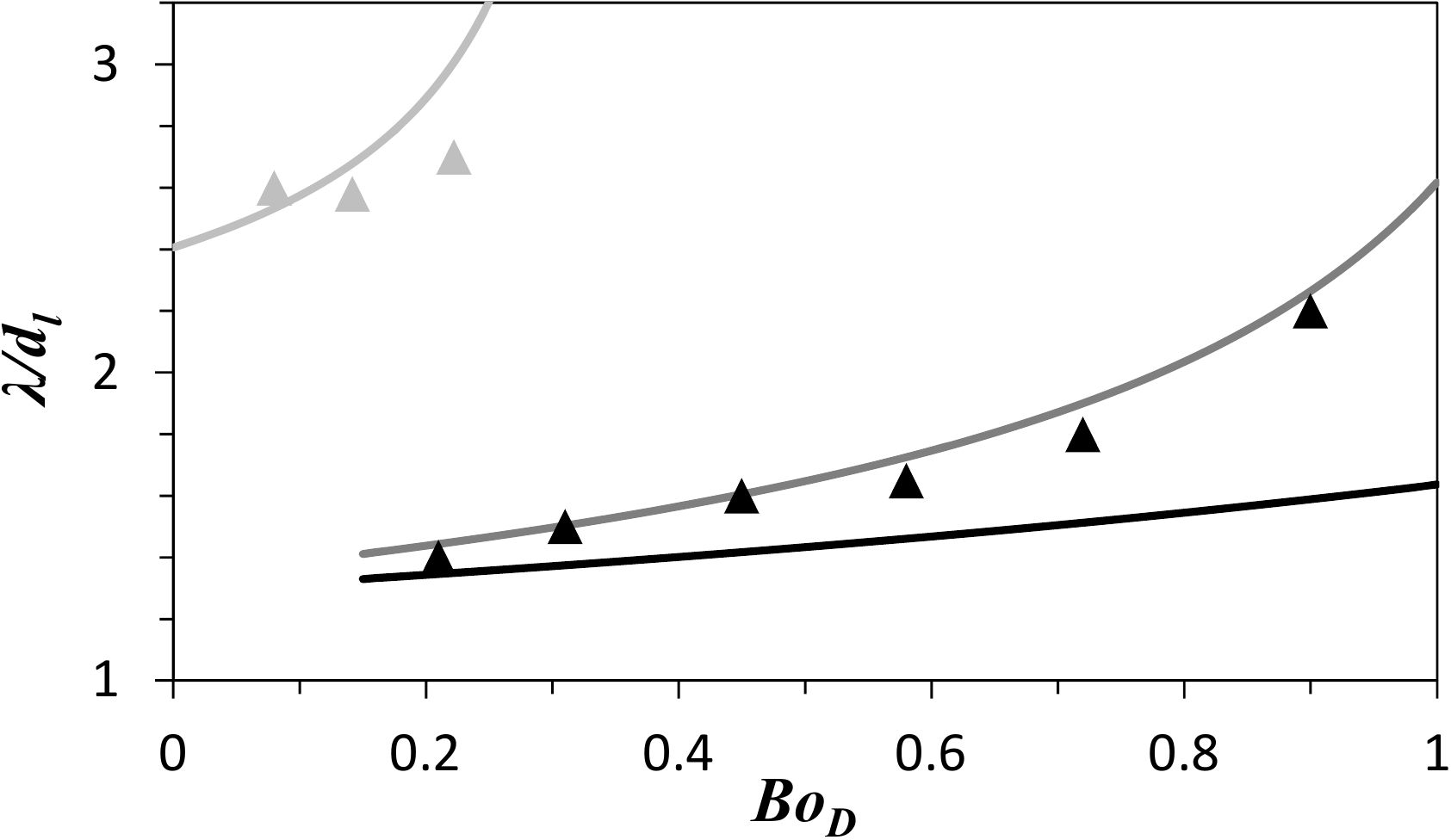}\vspace{-1mm}}
\caption{{Dependence of the critical Marangoni number (a) and the critical wavelength (b) on the dynamic Bond number for the 1.0 cSt silicone oil at atmospheric conditions ($c_a^0=0.995$) in a layer with $L_x=30$ mm. Experimental results obtained by \citet{riley1998} are shown as symbols (black filled for PMC/SMC, light gray filled for HTW, open for OMC). The boundary between HTW and OMC is shown as the dotted line. 
In both panels, analytical predictions of the two-sided model are shown as solid lines (light gray/dark gray/black for absolute instability/SMC/PMC threshold).}}
\label{fig:Ma-RN}
\end{figure}

{
The predictions of the linear stability analysis for this particular fluid are shown in Figs. \ref{fig:Ma-RN-inf} and \ref{fig:Ma-RN}.
The enhanced one-sided model and the two-sided model produce essentially identical results at atmospheric conditions; we only show the results of the latter.
In particular, the thresholds of different types of instabilities for a laterally unbounded system ($\Gamma_x=\infty$) are shown as solid lines in Fig. \ref{fig:Ma-RN-inf}(a). 
These are in decent agreement with the thresholds of convective, absolute, and stationary instabilities predicted by \citet{Priede1997} using an adiabatic one-sided model (which is equivalent to our enhanced one-sided model with $Bi_q=0$).
As expected, convective instability is predicted to occur at the lowest value of $Ma$ regardless of $Bo_D$.
Fig. \ref{fig:Ma-RN-inf}(b) shows the group velocity $v_g$ associated with convective instability (along with the phase velocity $v_p$ associated with absolute instability).
In the absence of end walls, one should expect a pattern arising from convective instability to develop sooner on one side than the other, with the interface between patterned and unpatterned flow propagating in the direction controlled by the sign of $v_g$, i.e., in the direction of the thermal gradient for $Bo_D<0.18$ and opposite the thermal gradient for $Bo_D>0.18$.}

{
In practice, the aspect ratio $\Gamma_x$ is too low for convective instability to amplify tiny disturbances that are naturally present in the flow sufficiently so they could be observed before they collide with one of the walls and disappear.
For instance, in the study of \citet{riley1998} $12<\Gamma_x<40$, while in most other experimental and numerical studies $\Gamma_x\le20$.
As a result, convective instability would not be observed (and will not be considered in detail here).
Instead an absolute instability, a stationary instability, or a combination of both instabilities would occur, and the order in which these instabilities occur would determine the type of convection pattern.
In particular, hydrothermal waves (HTW) are usually found for $Bo_D\to 0$ and a stationary pattern is found for $Bo_D=O(1)$.
However, we are not aware of any studies that predict what the pattern might look like at the intermediate values of $Bo_D$.
The analysis for large, but finite $\Gamma_x$ (so there are end walls) is presented below.
}

{
Comparison of the theoretical predictions with the results of \citet{riley1998} is complicated by the fact that in the experimental study the transition threshold for SMC is identified as the instant when multiple convection rolls appear (near the hot end wall), but do not extend all the way to the cold end wall, which is not a very precise definition.
However, the experimental data, which corresponds to an unknown value $-1<s<0$, should lie between the analytical prediction for the onset of PMC ($s=-1$) and SMC ($s=0$).
As shown in Fig.~\ref{fig:Ma-RN}, both the critical $Ma$ and the critical wavelength $\lambda$ observed in the experiments are indeed bracketed by the theoretical values corresponding to the onset of PMC and SMC, and lie closer to the SMC boundary in the entire range of $Bo_D$ investigated in the experiment, where stationary convective patterns are observed. 
Riley and Neitzel's supporting figures show the presence of numerous convection rolls at what they define to be the critical $Ma$, so their data points should indeed be closer to the SMC threshold then the PMC threshold.
}

{
For $Bo_D<0.04$ the absolute instability boundary lies below that of PMC, $Ma_c^a<Ma_c^s(-1)$, so a pattern should appear at $Ma_c^a$ over the entire extent of the fluid layer and feature traveling convection rolls, since $\omega_c\ne 0$.
The predicted phase velocity is positive, which is consistent with hydrothermal waves that are known to propagate from the cold end wall to the hot end wall.
All of this is in perfect agreement with the analysis of \citet{Smith1983a} applicable in the limit $Bo_D\to 0$.
Riley and Neitzel's experimental observations for $0.04<Bo_D<0.29$ deserve a separate discussion, since this is the range where the threshold of absolute instability lies between the PMC and SMC thresholds, $Ma_c^s(-1)>Ma_c^a>Ma_c^s(0)$.
In this range the base flow can undergo an instability with respect to two different patterns in different regions.
Consider the experimental protocol in which $Ma$ is gradually increased at a fixed $Bo_D$.
Our analysis predicts that a stationary pattern first develops near the hot end wall at the SMC threshold $Ma_c^s(-1)$.
As $Ma$ is increased beyond this value, the pattern spreads towards the cold end wall until $Ma$ reaches the threshold $Ma_c^a$ of absolute instability, which should cause hydrothermal waves to appear near the cold end wall.
Hence the fraction of the fluid layer occupied by the two patterns should then depend on $Bo_D$ and $\Gamma_x$, with HTW dominating for smaller $Bo_D$ and PMC at higher $Bo_D$, which is qualitatively consistent with experimental observations.
As Fig. \ref{fig:Ma-RN}(a) illustrates, our quantitative predictions for $Ma_c$ and $\lambda$ are also consistent with the experimental results in this range of $Bo_D$.
}

\begin{figure}
\subfigure[]{\includegraphics*[width=0.49\columnwidth]{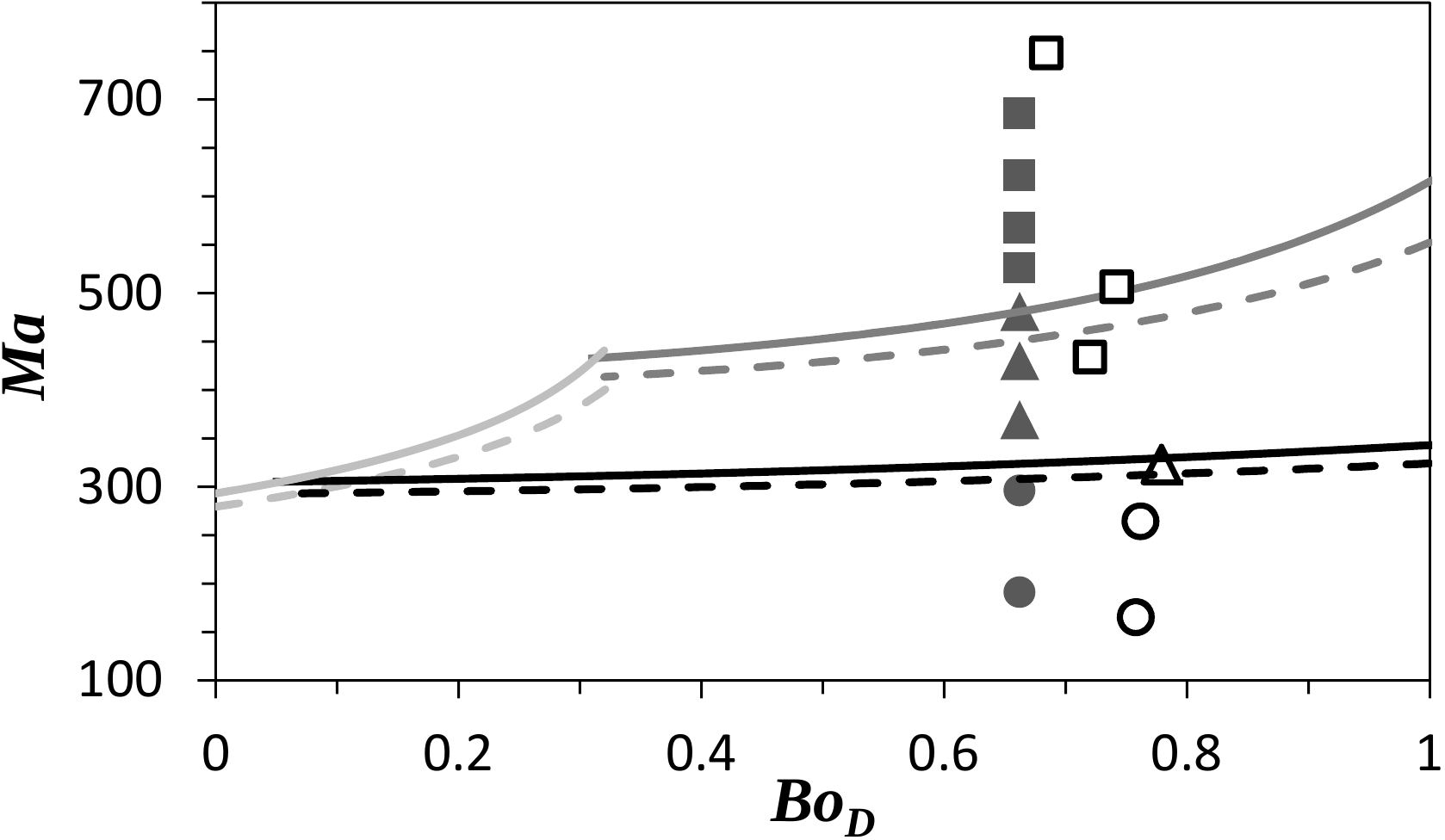}}\hspace{1mm}
\subfigure[]{\includegraphics*[width=0.49\columnwidth]{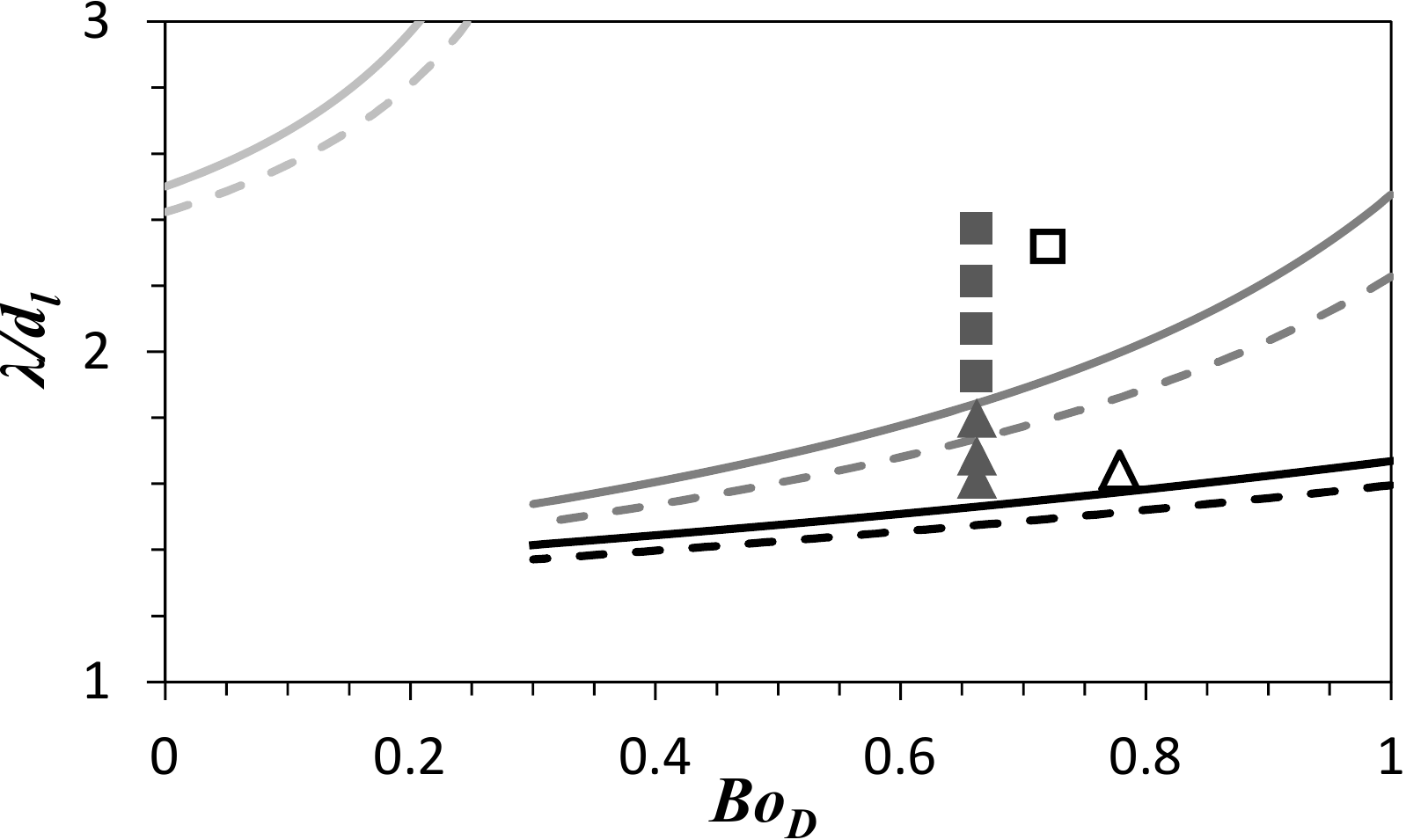}}
\caption{{Dependence of the critical Marangoni number (a) and the critical wavelength (b) on the dynamic Bond number for the 0.65 cSt silicone oil at atmospheric conditions ($c_a^0=0.96$). Open symbols correspond to experimental results of \citet{Li2013} and filled symbols -- to numerical results of \citet{Qin2013}. The four distinct flow regimes are: SUF ($\ocircle$), PMC ($\triangle$), SMC ($\Box$), and OMC ($\Diamond$). In both panels, analytical predictions of the two-sided model are shown as solid lines (light gray/dark gray/black for absolute instability/SMC/PMC threshold). Predictions of the one-sided model with $Bi_q=0$ are shown as dashed lines.}
}
\label{fig:Ma-Bo}
\end{figure}

{
If $Ma$ is increased further beyond $Ma_c^s(0)$, the SMC pattern itself eventually becomes unstable and is replaced by the oscillatory multicellular convection (OMC) state \citep{Villers1992,Desaedeleer1996,Garcimartin1997,riley1998,Qin2013}.
OMC features convection rolls traveling in the direction opposite to that of HTW, i.e., against thermal gradient, with the right-most convection roll acting as the oscillatory source of disturbance \citep{Garcimartin1997,Qin2013,Li2013}.
While it may be tempting to associate the extension of the curve $Ma=Ma_c^a$ for $Bo_D>0.29$ in Fig. \ref{fig:Ma-RN}(a) with the boundary between SMC and OMC, it is incorrect to do so, since the absolute instability describes destabilization of the uniform base flow rather than the SMC pattern.
On the other hand, the group velocity $v_g$ associated with convective instability appears to predict the location of the boundary between HTW and OMC (the dashed line in Fig. \ref{fig:Ma-RN}(a)).
As mentioned previously, $v_g>0$ for $Bo_D<0.18$, such that disturbances originating near the cold end wall should amplify as they propagate towards the hot end wall, while $v_g<0$ for $Bo_D>0.18$, so the opposite is true, which is consistent with the location of the wave sources for HTW and OMC. For comparison, in the experiment the boundary is at $Bo_D=0.22$.
}

Next we compare the analytical predictions with the results of numerical and experimental studies \citep{Qin2013,Qin2015,Li2013}, which used a more volatile liquid (0.65 cSt silicone oil with $Pr=9.2$) and characterized both the SMC$\to$PMC and the PMC$\to$SMC transition. 
The predicted dependence {of the critical Marangoni number and wavelength} on the dynamic Bond number at atmospheric conditions ($c_a^0=0.96$) is shown in Fig.~\ref{fig:Ma-Bo}. 
Again, the predictions of the enhanced one-sided and two-sided model are essentially indistinguishable and are in reasonable agreement with both experimental and numerical data.
Just as in the case of 1 cSt silicone oil, the critical values of $Ma$ and $\lambda$ increase monotonically with $Bo_D$ over the range where a stationary pattern is found {($Bo_D>0.31$)}. 
This increase in $Ma_c$ can be easily understood by noting that buoyancy has a stabilizing effect on the flow.
The base flow solution \eqref{eq:Ttheo} shows that the temperature increases, and hence the density of the liquid decreases, with height. 
An increase in $Bo_D$ therefore corresponds to an increasingly strong effect of the stable density stratification. 

{
At low $Bo_D$ we again find that the absolute instability precedes the stationary instability, and we should expect to find HTW, at least near the cold end wall.
According to Fig. \ref{fig:Ma-Bo}(a), for $Bo_D<0.05$, the threshold of absolute instability lies below the PMC threshold, $Ma_c^a<Ma_c^s(-1)$, so HTW should be observed over nearly the entire extent of the fluid layer.
For $0.05<Bo_D<0.31$, the threshold of absolute instability lies between the PMC and SMC thresholds, $Ma_c^s(-1)<Ma_c^a<Ma_c^s(0)$.
As we discussed previously, in this range we expect the two patterns to coexist for sufficiently large $\Gamma_x$, with the boundary between HTW and PMC at onset of the absolute instability gradually shifting towards the cold end wall as $Bo_D$ increases.
The competition between these patterns above the threshold should be described by nonlinear theory and hence is not addressed here.
}

\begin{figure}
\subfigure[]{\includegraphics*[width=0.49\columnwidth]{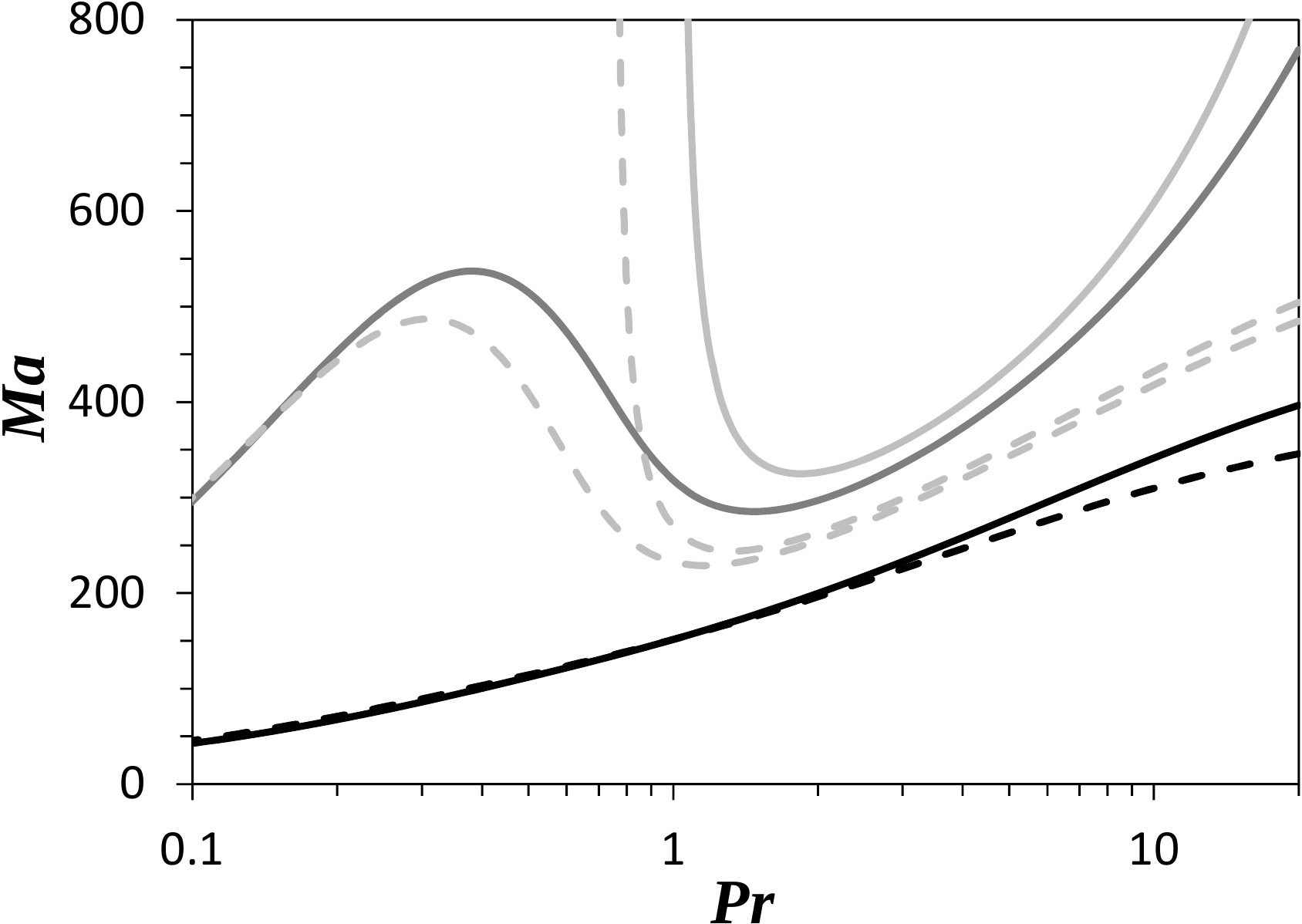}}\hspace{1mm}
\subfigure[]{\includegraphics*[width=0.49\columnwidth]{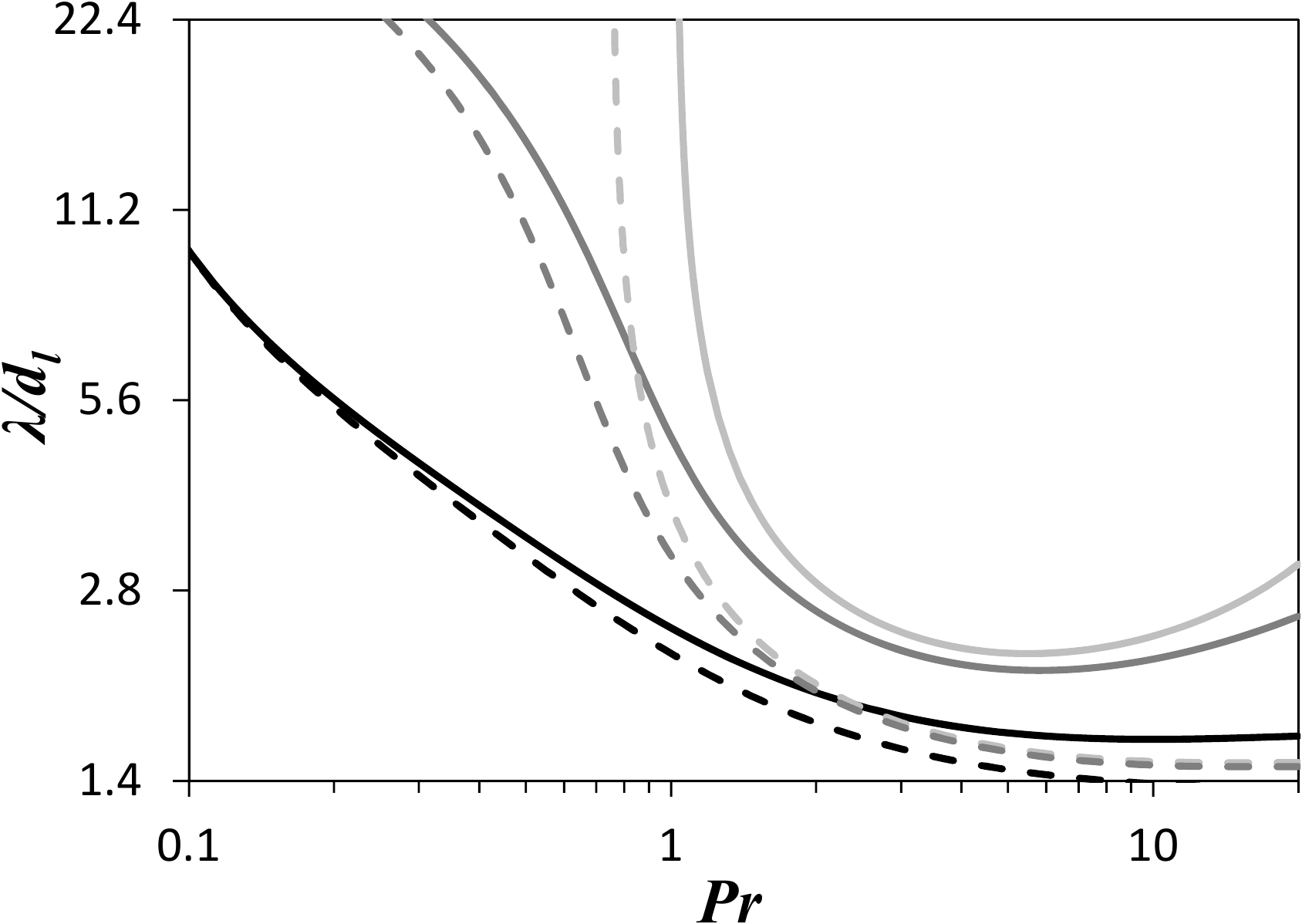}}
\caption{{Stationary instability. Dependence of the critical Marangoni number (a) and the critical wavelength (b) on the Prandtl number of the liquid for $K=0.2$ and $c_a^0\to 1$. The lines (solid for $Bo_D=1$, dashed for $Bo_D=0.3$) show the prediction of the enhanced one-sided model with $s = -1$ (black), $s = -0.1$ (dark gray), and $s = 0$ (light gray). }
}
\label{fig:Ma-Pr}
\end{figure}

\begin{figure}
\subfigure[]{\includegraphics*[width=0.49\columnwidth]{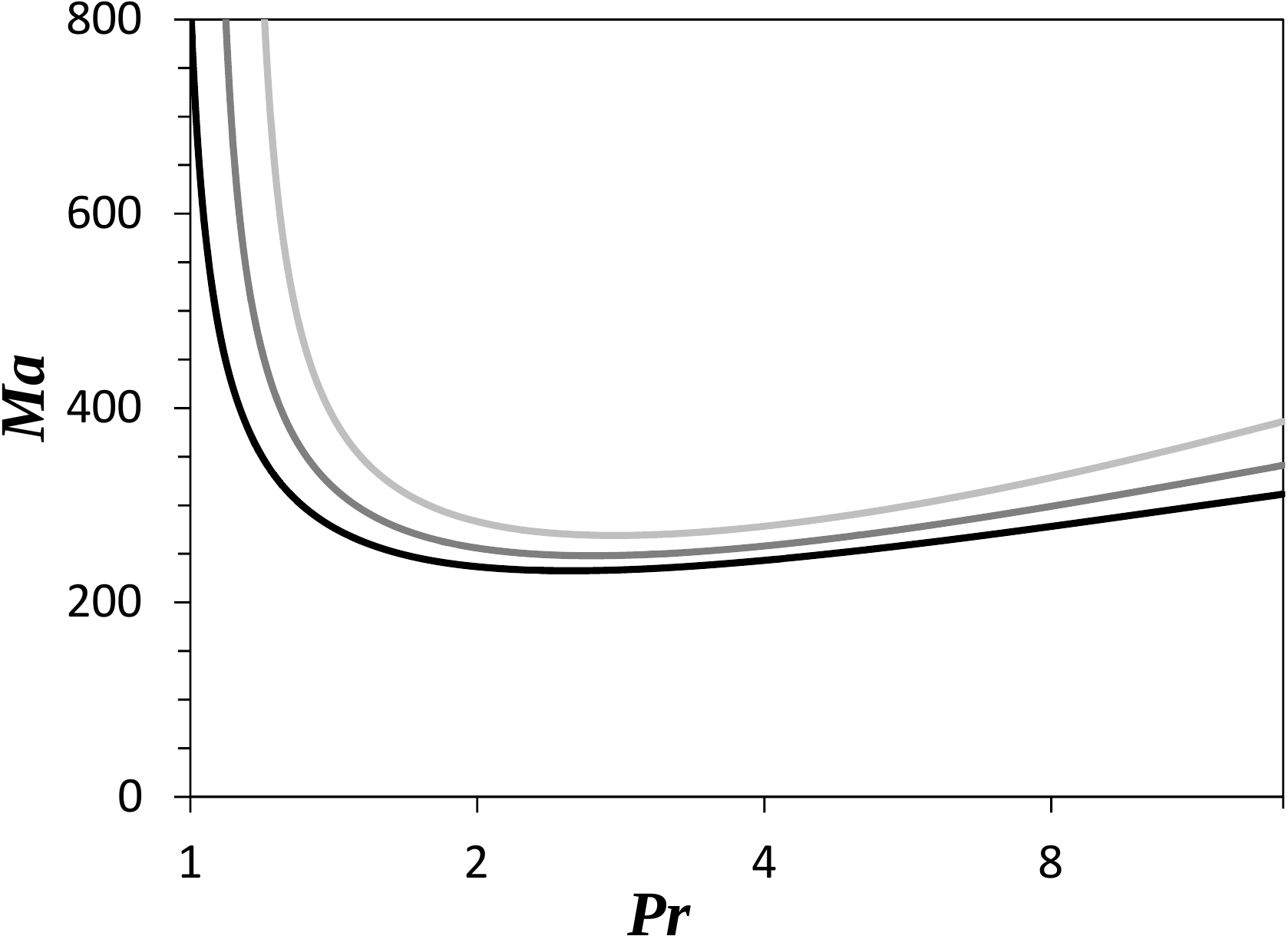}}\hspace{1mm}
\subfigure[]{\includegraphics*[width=0.49\columnwidth]{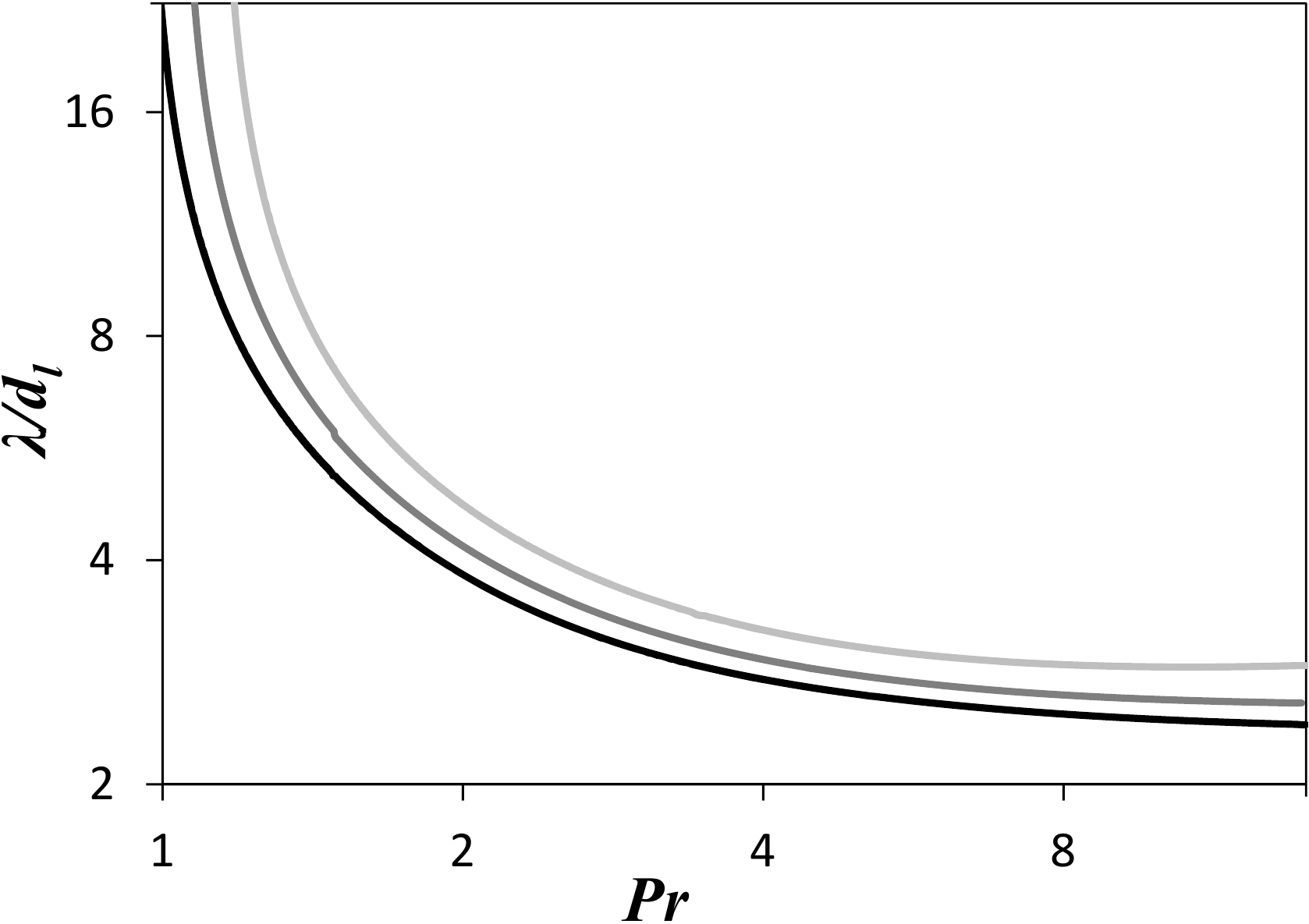}}
\caption{{Absolute instability. Dependence of the critical Marangoni number (a) and wavelength (b) on the Prandtl number of the liquid for $K=0.2$ and $c_a^0\to 1$. The lines show the prediction of the enhanced one-sided model for $Bo_D = 0$ (black), $Bo_D = 0.1$ (dark gray), and $Bo_D = 0.2$ (light gray).}
}
\label{fig:abs_Pr}
\end{figure}

The dependence of the threshold values of $Ma$ and $\lambda$ on the Prandtl number of the liquid has previously been investigated only in the context of linear stability analysis. 
In particular, \cite{Priede1997}, who only considered uniform disturbances ($s=0$) and ignored heat and mass flux across the interface, found that the critical values of $Ma$ for the stationary instability diverges at $Pr_0=O(1)$ {and reaches a minimum at $Pr_1$ just slightly higher than $Pr_0$. 
(Oddly enough, they only show the data for $Bo_D=0$, where HTW are expected instead of SMC.)
Our enhanced one-sided model predicts that this result also holds for $Bo_D=O(1)$, once transport in the gas phase has been taken into account (cf. Fig.~\ref{fig:Ma-Pr}).
The corresponding data for the absolute instability at lower $Bo_D$ is shown in Fig. \ref{fig:abs_Pr}.
Predictions of the two-sided model are not shown, since it involves too many parameters which will vary with $Pr$, as different $Pr$ correspond to different liquids. }

{Qualitatively similar trends are found for the threshold of SMC for all $Bo_D=O(1)$, with both $Pr_0$ and $Pr_1$ slowly increasing with $Bo_D$. }
Over the entire range of $Pr$ corresponding to nonmetallic liquids, $Ma_c$ is a monotonically increasing function {of $Pr$}.
The dependence of the critical wavelength  on the Prandtl number is {nonmonotonic even for nonmetallic liquids}; $\lambda$ has a minimum at {$Pr_2=O(10)$ and also diverges at $Pr_0$.
The divergence of the critical values of $Ma$ and $\lambda$ at $Pr_0$, however, does} not imply that no stationary pattern emerges at $Pr<Pr_0$. 
Indeed, the critical $Ma$ for PMC ($s=-1$) decreases monotonically as $Pr$ decreases, with $Ma_c\to 0$ as $Pr\to 0$. 
{For instance, for a moderate value of $Ma\approx 600$, a stationary (spatially nonuniform) pattern is predicted to appear near the hot end wall for {\it any} $Pr<Pr_0$.} 
As $Ma$ increases, the spatial extent of the pattern should increase as well. 

\begin{figure}
\subfigure[]{\includegraphics*[width=0.5\columnwidth]{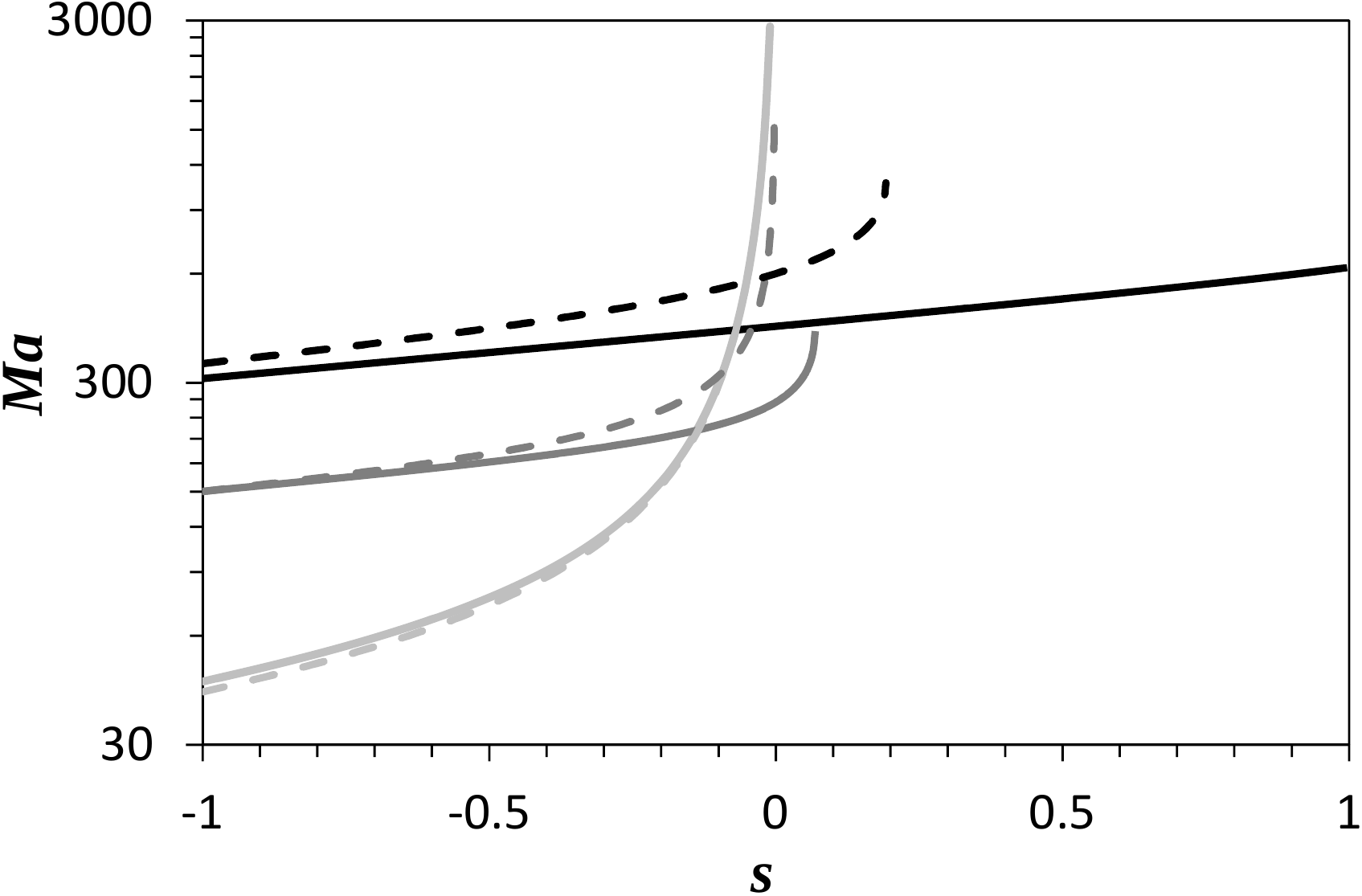}}\hspace{1mm}
\subfigure[]{\includegraphics*[width=0.49\columnwidth]{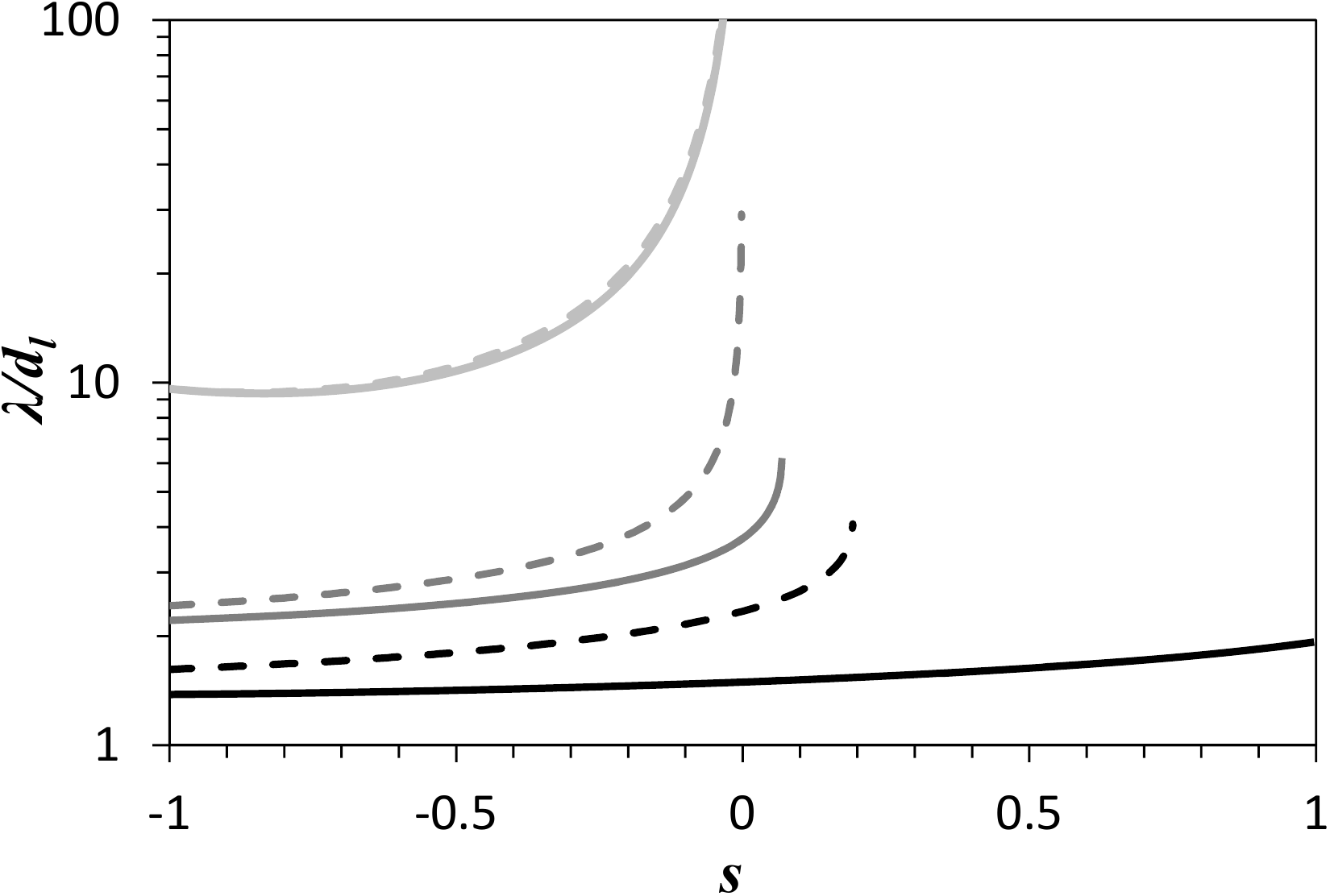}}
\caption{{Stationary instability. Dependence of the critical Marangoni number (a) and wavelength (b) on the spatial attenuation rate for $c_a^0\to 1$, $K=0.2$, and $Pr=10$ (black), $Pr=1$ (dark gray), or $Pr=0.1$ (light gray). Solid lines correspond to the predictions of the enhanced one-sided model with $Bo_D=0.3$ and dashed lines to $Bo_D=1$. }
}
\label{fig:Re-ki}
\end{figure}

As noted in the introduction, the analysis of \citet{Mercier2002} predicted that, steady convection rolls should develop near the hot end for $Pr>4$, near the cold end for $Pr<0.01$, or at both end walls for $0.01<Pr<4$. 
Our results unequivocally contradict that prediction {for $Bo_D=O(1)$. 
According to Fig. \ref{fig:Ma-Pr}(b), for $Pr\lesssim 0.1$ the wavelength of the pattern exceeds typical values of $\Gamma_x$ for any value of the the spatial attenuation rate $s$, so it makes no sense to discuss spatial localization.
The dependence of the critical $Ma$ and $\lambda$ on $s$ for the stationary instability at $Pr=0.1$, 1, and 10 are shown in Fig. \ref{fig:Re-ki}). 
We find that $Ma_c^s$ is a monotonically increasing function of $s$, regardless of the values of $Pr$ for $Bo_D=O(1)$, indicating that a stationary pattern {\it always} emerges first near the hot end wall.}

{The analysis of \citet{Mercier2002} was performed mainly to understand the results of numerical simulations by \citet{BenHadid1990} in the limit of low $Pr$ which indicated the formation of a convective pattern near the cold end wall. 
Both of these studies made two contradicting assumptions: (1) that the dynamic Bond number $Bo_D\ll 1$ and (2) that the liquid-gas interface is flat. 
As \citet{Smith1983b} showed, in the limit $Bo_D\to 0$ the uniform base flow undergoes an instability towards surface waves for $Pr<0.15$, violating the assumption of flat interface before a stationary instability can occur.
Indeed, as Fig. \ref{fig:abs_Pr}(b) shows, even for finite $Bo_D$ the critical wavelength associated with the absolute instability diverges as $Pr$ approaches $Pr_0=O(1)$ from above, indicating that for $Pr<Pr_0$ a long-wavelength instability will arise that necessarily causes substantial deformation of the interface.
Even for $Pr>Pr_0$, no stationary pattern will appear near the cold end wall for small $Bo_D$, since the absolute instability precedes the stationary one, as Figs. \ref{fig:Ma-RN} and \ref{fig:Ma-Bo} show. 
}

{In conclusion of this section, let us point out that even} though phase change is greatly suppressed at atmospheric conditions, this does not mean that the effect of {the gas phase is negligible. 
Fig. \ref{fig:Ma-RN-inf} clearly shows the difference in the thresholds of all three instability types predicted by the adiabatic one-sided model (which completely ignores the effect of the gas layer) and the two-sided model (which accounts for heat, mass, and momentum transport in the gas) for a fluid of relatively low volatility (1 cSt silicone oil).
The difference is even more significant for the more volatile 0.65 cSt silicone oil, as Fig. \ref{fig:Ma-Bo} illustrates.
We find that the one-sided model underestimates the critical values of both $Ma$ and $\lambda$ which correspond to the PMC/SMC boundary by between 4\% at $Bo_D\approx 0.3$ and 11\% at $Bo_D\approx 1$.
The corresponding discrepancy for the absolute stability boundary ranges from 3\% at $Bo_D=0$ to 10\% for $Bo_D\approx 0.3$.
Not surprisingly, we find that transport of heat and mass through the gas phase delays all of the transitions. 
The effect of the gas phase} increases rather dramatically, however, once the pressure is reduced below the atmospheric value, as we show below.

\subsection{Convection at reduced pressures}

\begin{figure}
\subfigure[]{\includegraphics*[width=0.49\columnwidth]{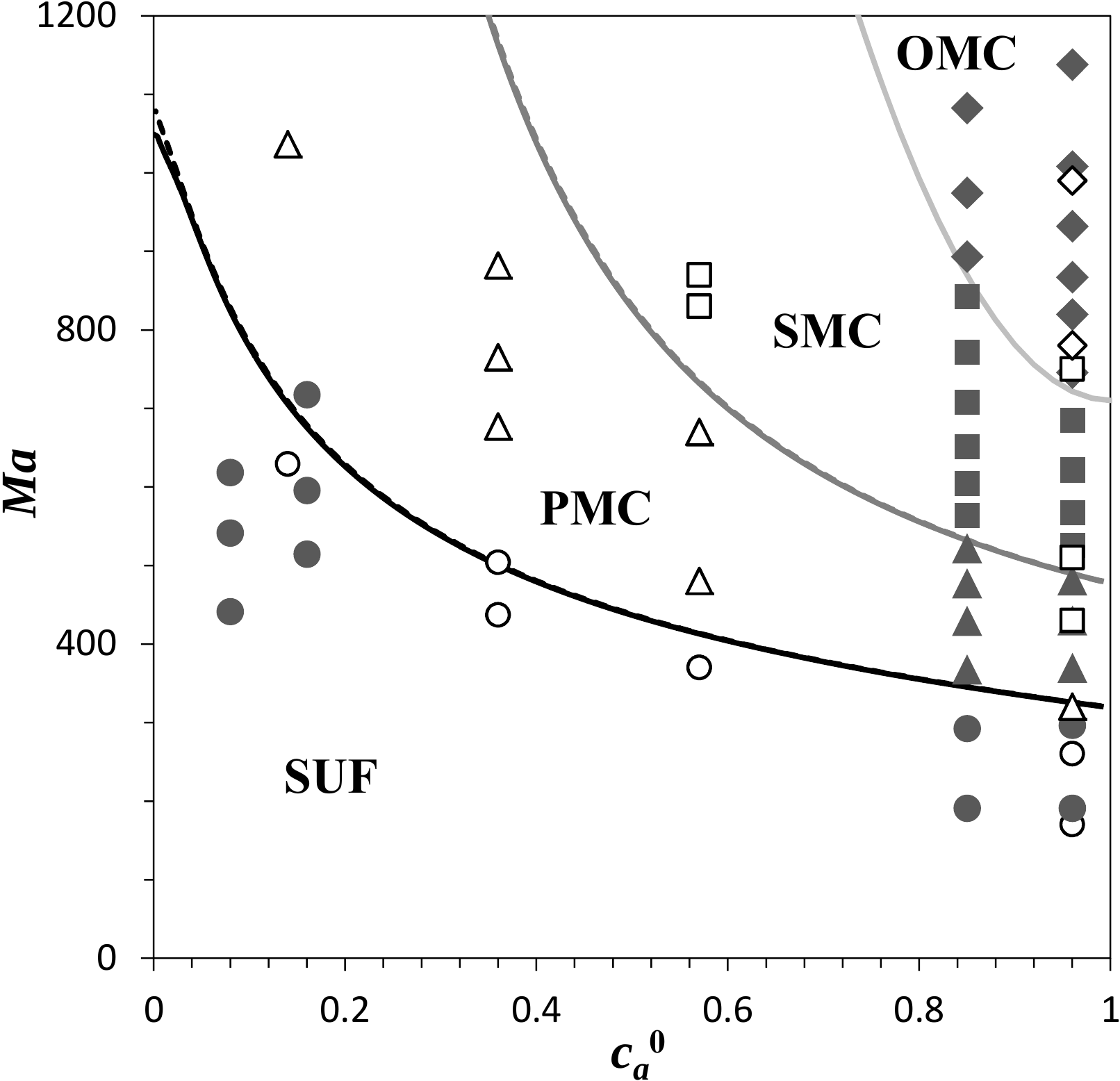}}\hspace{1mm}
\subfigure[]{\includegraphics*[width=0.49\columnwidth]{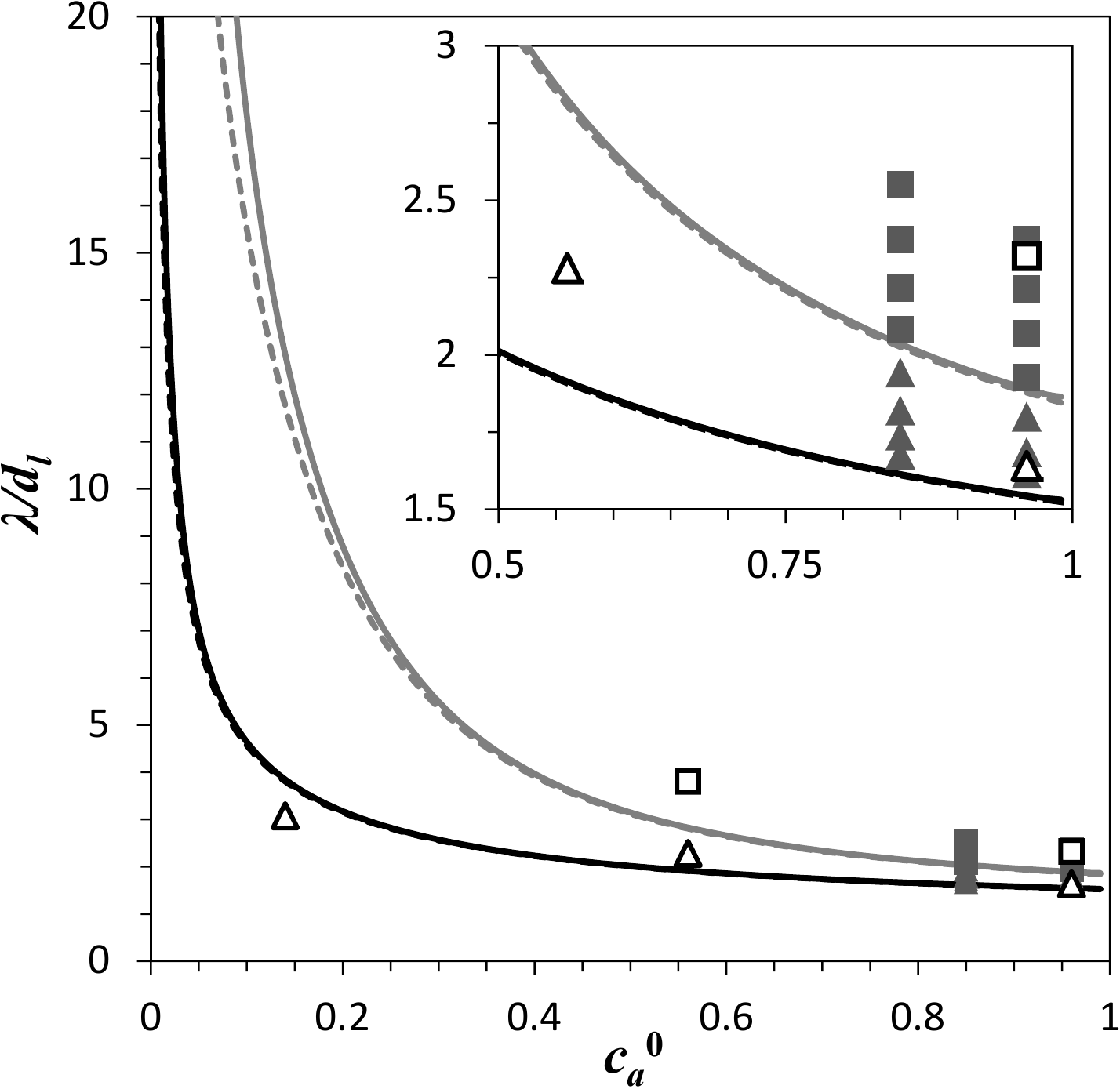}}
\caption[]{{Dependence of the critical Marangoni number (a) and wavelength (b) on the average concentration of air for the 0.65 cSt silicone oil with $Bo_D=0.7$. Solid (and dashed) lines are predictions of the two-sided (and enhanced one-sided) model for the thresholds of the PMC (black), SMC (dark gray), and OMC (light gray) patterns. The OMC threshold is a sketch based on numerical and experimental data. The meaning of the symbols is the same as that in Fig.~\ref{fig:Ma-Bo}. The inset in panel (b) shows the region of high $c_a^0$ and low $\lambda$.}
}
\label{fig:Ma-ca}
\end{figure}

In this section we will discuss the dependence of the critical $Ma$ on the average concentration $c_a^0$ of air. We will focus exclusively on the 0.65 cSt silicone oil and {$Bo_D = 0.7$, 
since relevant numerical and experimental data tend to cluster around that value.
As we discussed previously, at this $Bo_D$ the stationary instability dominates, so a time-independent pattern emerges.
The one-sided model of \citet{Priede1997} does not account for phase change (recall that it corresponds to our enhanced one-sided model with $Bi_q=0$), so its predictions are independent of $c_a^0$.}
The predictions of the enhanced one-sided and the two-sided model are compared with available data in Fig.~\ref{fig:Ma-ca}. 
The results of the two models are essentially identical, except for low $c_a^0$, where the validity of the assumptions underlying both models is questionable, as discussed below.

The predicted $Ma_c$ increases rather significantly as $c_a^0$ is decreased from the atmospheric values down to the regime when the gas is dominated by the vapour, rather than air.
The results of linear stability analysis are {generally} consistent with the experimental data, although at atmospheric conditions the predicted thresholds for the onset of PMC and SMC states are higher by about 20\% compared with those found in experiment  (cf. Fig.~\ref{fig:Ma-ca}(a)).
Numerical simulations have only been performed in the limits when either air or vapour dominate.
When the gas phase is dominated by air ($c_a^0\ge 0.85$), {the predictions of linear stability analysis are consistent with the results of numerical simulations.}
When the vapour is dominant ($c_a^0 \le 0.16$), the numerical simulations only find the SUF flow over the limited range of {the imposed temperature gradients} considered, so no prediction of $Ma_c$ can be made. 
However, the {corresponding lower bound supports} the theoretical prediction {that the threshold increases considerably}.

The dependence of the critical wavelength on $c_a^0$ shows {a trend similar to $Ma_c$: the predicted $\lambda$ increases as $c_a^0$ decreases (cf. Fig.~\ref{fig:Ma-ca}(b)).}
These predictions are in reasonable agreement with the available numerical and experimental results \citep{Qin2015,Li2013}. 
When the gas phase is dominated by air ($c_a^0\ge 0.85$), the {predicted critical wavelength for both the PMC and SMC state is in good agreement with} the values found in the numerics {and experiment}.
Only experimental data is available for lower $c_a^0$, and again it is consistent with theoretical predictions for $c_a^0>0.3$: the wavelength of the pattern is found to lie between the PMC and SMC curves, as it should.
{For $c_a^0< 0.16$ the critical wavelength becomes comparable to the length $L$ of the cavity studied in \citet{Li2013}, so some discrepancy between the theoretical predictions and the experimental results is expected.}

{In general, our linear stability analysis assumes a flat interface and will break down for $\lambda\gg d_l$ (i.e., at low $c_a^0$ and/or low $Pr$), when the thickness of the liquid layer is expected to vary noticeably. }
Furthermore, for sufficiently low $c_a^0$ and/or {large $\Gamma_x$} the interfacial temperature profile in the SUF state deviates significantly from linear, resulting in the breakdown of the analytical solution for the base flow \citep{Qin2015}.  
Therefore, the theoretical predictions {in these limits} are not expected to be particularly accurate.

\section{Discussion}
\label{sec:dis}

As our results illustrate, the predictions of the linear stability analysis based on the two-sided model are in general agreement with the reported numerical and experimental data. In particular, this analysis correctly predicts that the transitions between different flow regimes (SUF, PMC, SMC) are delayed (i.e., shifted towards higher $Ma$) when the concentration of air decreases, although quantitative agreement is not expected at low $c_a^0$ when the assumptions {underlying our analysis become} invalid. The general agreement found at higher $c_a^0$ confirms the validity of the assumptions made in the linear stability analysis, giving us confidence that it captures all the essential physical processes governing the flow stability for volatile liquids driven by a horizontal temperature gradient.

{
The predictions of linear stability analysis are consistent with the available numerical results.}
The discrepancy with the experiment can be due to a number of different reasons. One is the effect of transverse confinement: the liquid is wetting and rises at the side walls, increasing the thickness of the liquid layer there by up to 60\%. The experimental study by \citet{Li2013} used the (smaller) thickness $d_l$ of the liquid layer at the {symmetry mid-plane $y=L_y/2$} of the cavity to compute $Ma$. Furthermore, the experimental study indirectly deduced the interfacial temperature gradient {by comparing the experimental velocity profile with the analytical solution as in Fig. \ref{fig:an_vs_num}(a)}
rather than measure it directly, which introduces additional uncertainty into the reported values of {$Ma_c$}.

In this problem, buoyancy plays a stabilizing role, hence the instability leading to the formation of convection rolls is driven primarily by thermocapillary stresses, which depend on the interfacial temperature gradient. 
The fluctuations in the interfacial temperature are significantly affected by the heat and mass transport through the gas phase {described -- in the enhanced one-sided model --} by the Biot coefficient $Bi_q$, which is a function of both the wavenumber $q$ and the concentration of air $c_a^0$. 
{In particular, the} first term in the square brackets in (\ref{eq:Bi2}) 
\begin{align}
{B_1=\frac{k_g}{k_l}=\frac{c_a^0k_a+(1-c_a^0)k_v}{k_l}}
\end{align}
describes the effect of conductive heat transport through the gas layer and reflects the dependence of thermal conductivity of the gas phase on {its} composition, while the second term 
\begin{align}
{B_2=\frac{1-c_a^0}{c_a^0}H},
\end{align}
where $H$ is the nondimensional parameter defined by \eqref{eq:H}, describes the effect of the latent heat released or absorbed at the interface as a result of phase change and reflects the variation in the diffusive transport of vapour through the gas phase, which also depends on the gas composition.
Both heat and mass transport through the gas phase suppress fluctuations in the interfacial temperature {and hence play a stabilizing role}.

\begin{figure}
\subfigure[]{\includegraphics*[width=0.49\columnwidth]{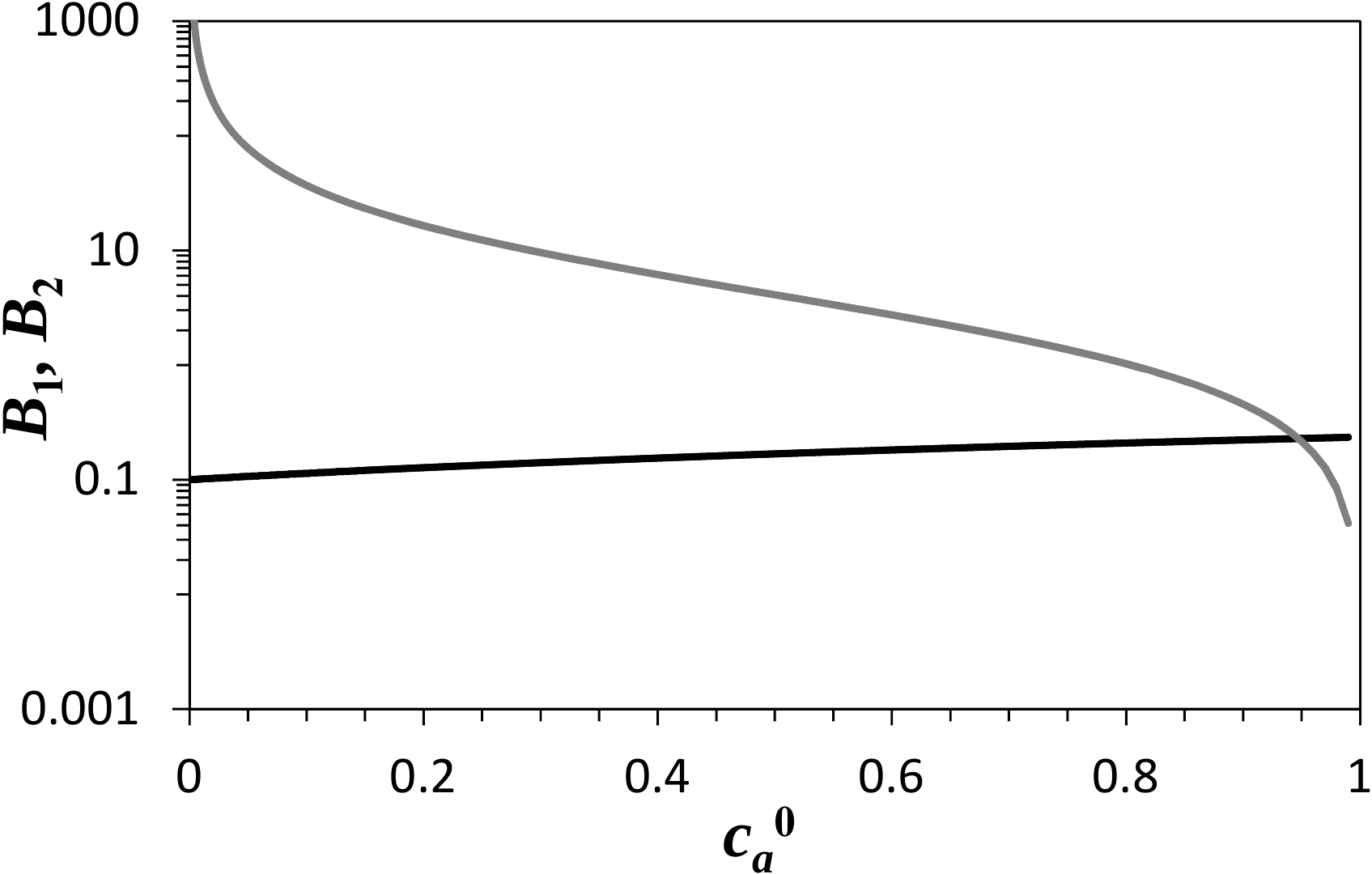}}\hspace{1mm}
\subfigure[]{\includegraphics*[width=0.49\columnwidth]{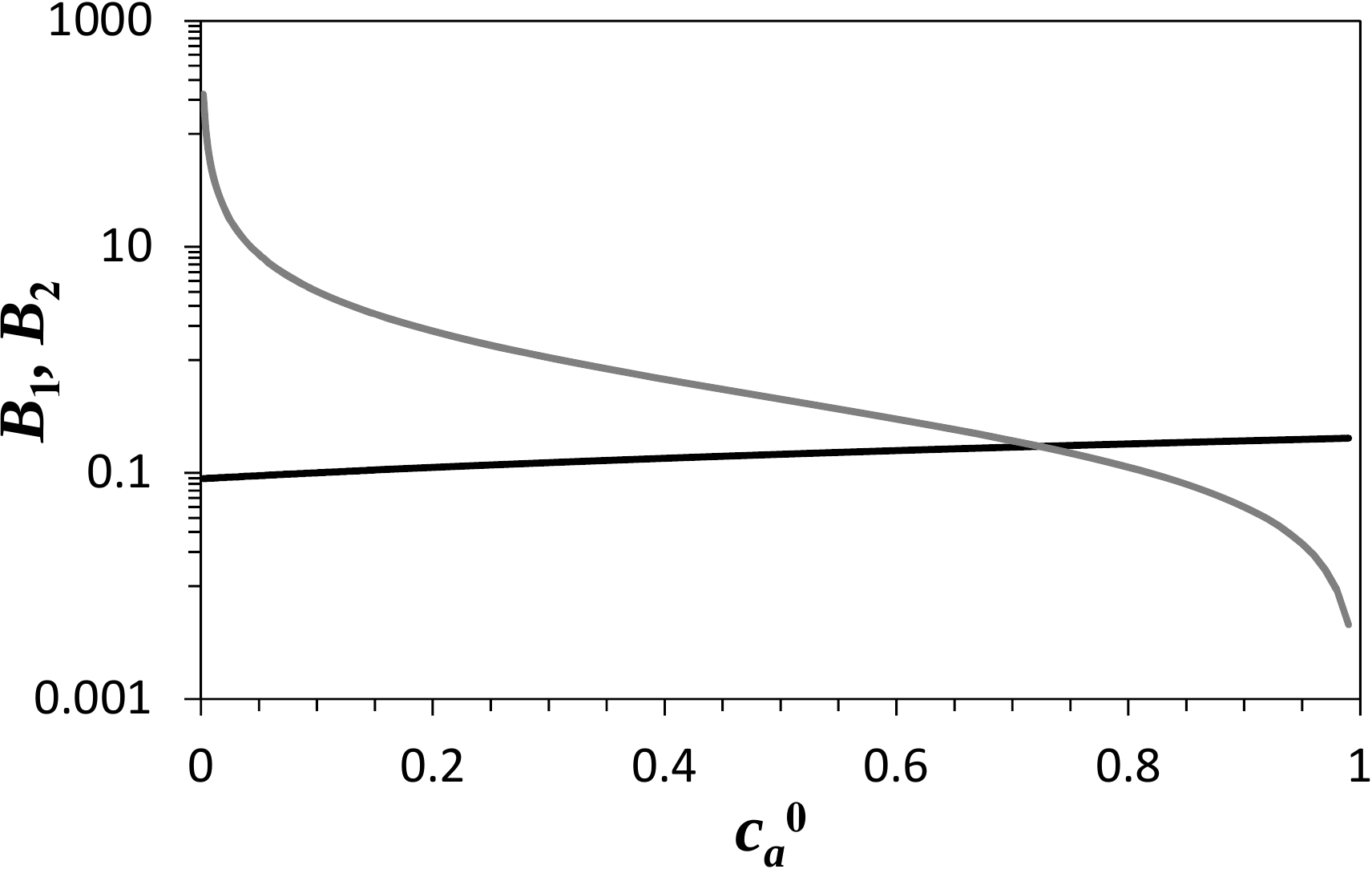}}
\caption[]{{The role of heat and mass transport in the gas layer on flow stability for (a) the 0.65 cSt silicone oil and (b) the 1 cSt silicone oil. $B_1$ is shown as black and $B_2$ as gray line.}
}
\label{fig:Biq}
\end{figure}

To see the relative impact of the heat and mass transport on stability, it is instructive to compare the trends and characteristic magnitudes of the two terms. 
{For the 0.65 cSt silicone oil, $B_1$ decreases slightly as $c_a^0$ decreases (thermal conductivity of the vapour is somewhat smaller than that of the air), but $B_2$ increases significantly, reflecting the enhancement of phase change (cf. Fig. \ref{fig:Biq}). At atmospheric conditions, the heat flux in the gas phase has a slightly larger effect ($B_1=0.23$ vs. $B_2=0.17$). The effect of phase change becomes dominant for $c_a^0<0.95$ and increases as $c_a^0$ decreases. In fact, $B_2$ diverges as $c_a^0\rightarrow 0$ which, according to (\ref{eq:qper2}), implies that the temperature fluctuations at the interface vanish and the critical $Ma$ becomes infinite. The trends are similar for the less volatile 1 cSt silicone oil, although the effect of phase change is clearly weaker.}

The increase in the critical $Ma$ with decreasing $c_a^0$ is due primarily to the enhancement of phase change which increases the amount of latent heat released/absorbed by the warmer/cooler regions of the interface.
Interestingly, the composition of the gas phase has a more significant effect on flow stability than the base flow itself. 
As discussed by \citet{Qin2015}, the concentration of air has a relatively weak effect on the base flow, since the average interfacial temperature gradient, which determines the thermocapillary stresses and hence the speed of the base flow, is insensitive to $c_a^0$ above 10\% or so. 

{The one-sided model with the adiabatic boundary condition is ill-suited for describing the flow of volatile fluids. 
As discussed previously, even at atmospheric conditions, when phase change is strongly suppressed, 
it underestimates the critical $Ma$ corresponding to all three instability types.
At reduced pressures, phase change at the interface and heat and mass transport through the gas phase are all significantly enhanced, and predictions of the one-sided model become very inaccurate, since it completely fails to describe these effects.
To illustrate the role of the gas phase,} it is useful to focus on the dimensionless combinations that have been introduced over the years to describe the deviation from thermodynamic equilibrium, which gives rise to convection in the presence of evaporation. In particular, in their analysis of pure Marangoni (thermocapillary) instability in volatile fluids subjected to a {\it vertical} temperature gradient, \citet{Burelbach1988} introduced two nondimensional parameters: the evaporation number
\begin{align}\label{eq:E}
E=\frac{k_l\Delta T}{\rho_l\nu_l\mathcal{L}},
\end{align}
which defines the ratio of the evaporative time scale (how long it would take for a liquid layer to evaporate completely) to the viscous time scale, and the ``nonequilibrium parameter'' \begin{equation}
\label{eq:K}
K=\frac{2\chi}{2-\chi}\sqrt{\frac{\bar{R}_vT_0}{2\pi}}
\frac{\rho_v\mathcal{L}^2d_l}{k_l\bar{R}_vT_0^2},
\end{equation}
which defines the ratio of the latent heat flux at the interface to the conductive heat flux in the liquid. 
The stability analysis of that problem was performed correctly only recently by \citet{Chauvet2012} who found that a version of the Biot coefficient (\ref{eq:Bi2}) appears there as well. 
However, neither $E$ nor $K$ appears in the stability analysis, regardless of the direction of the temperature gradient, because both parameters fail to account for the transport (of either heat or mass) in the gas phase.
Instead, the Biot coefficient involves a nondimensional combination $H$ which describes the effect of the latent heat associated with phase change and explicitly takes into account the mass transport in the gas phase.

The study by \citet{Normand1977} was probably the first to note that the Biot ``number'' should depend on the wavenumber {of the pattern} for convective flows. In their analysis of buoyancy-thermocapillary convection driven by a horizontal temperature gradient, \citet{Mercier1996} repeated that statement, but proposed to use the definition
\begin{align}\label{eq:Bic}
Bi=\frac{k_gd_l}{k_ld_g}
\end{align}
based on conduction in the two layers instead. Although wavenumber-independent, this definition is similar in form to the correct one in the {combined limits of an infinitely thick gas layer and vanishing volatility.} Indeed, if we ignore phase change, for $d_g\gg d_l$, the expression (\ref{eq:Bi2}) would reduce to 
\begin{align}
Bi_q=q\frac{k_g}{k_l}=2\pi\frac{k_gd_l}{k_l\lambda}.
\end{align}
Here the wavelength of the pattern effectively plays the role of the gas layer thickness, the result that follows immediately from the second equation in \eqref{eq:diffg} which controls the heat transport in the gas phase.

The wavenumber dependence of the Biot coefficient (\ref{eq:Bi2}) implies that, once transformed to the real space, Newton's cooling law
\begin{align}\label{eq:NLC}
d_l\hat{\bf n}\cdot\nabla T_l=-Bi(T_i-T_0)
\end{align}
becomes invalid and has to be replaced by a spatially nonlocal relation between the interfacial temperature and the normal temperature gradient (or heat flux) in the liquid phase. Given that the Biot number \eqref{eq:Bic} describes one-dimensional conduction, it is not surprising that \eqref{eq:NLC} fails when spatial variation of the solution in the extended direction(s) is taken into account.

Finally, by comparing the results of linear stability analysis based on three models of the gas layer that incorporate varying levels of detail, we found that advection (of heat, mass, or momentum) plays a relatively minor role and can be ignored except for {low $c_a^0$ (e.g., when vapor dominates in the gas phase).
In this limit, the two-sided model, which accounts for advection in the gas layer predicts a slightly higher critical $Ma$ for the threshold of both PMC and SMC. 
The increased role of advection can be readily understood by considering the associated changes in the P\'eclet numbers \eqref{eq:Pem} and \eqref{eq:Pet}. 
Both $\alpha_g$ and $D$ are proportional to $p_g^{-1}\propto 1-c_a^0$, so both $Pe_m$ and $Pe_t$ should increase with decreasing $c_a^0$ (even if we ignore the increase in the characteristic velocity of the gas associated with enhanced phase change).
We should also expect advection to play a more important role for low $Pr$, since both $Pe_m$ and $Pe_t$ are proportional to $Re=Ma/Pr$. The enhanced advective transport in the gas phase at low $c_a^0$ or low $Pr$} tends to suppress the variation in the temperature at the interface, the thermocapillary stresses, and hence the onset of convection.

\section{Conclusions}\label{sec:sum}

As several recent experimental \citep{Li2013} and numerical \citep{Qin2015} studies have demonstrated, phase change has a rather dramatic effect on the stability of the flow in a layer of  a volatile liquid (0.65 cSt silicone oil) driven by a horizontal temperature gradient. These studies showed that the onset of instability is {delayed} rather significantly when the phase change is enhanced by removing noncondensable gases, such as air, from the system. Linear stability analysis {based on a model that takes heat and mass transport in the gas phase into account} produces results that are overall in good quantitative agreement with these studies. In particular, it confirms that the instability that is responsible for generating a pattern of convection rolls is caused by the thermocapillary stresses, while buoyancy in the liquid layer has a stabilizing effect.

The analysis also shows that {phase change plays a very important role, with} the magnitude of the effect described by the nondimensional combination $H$ defined by \eqref{eq:H}, the same combination that appears in the convection problem with a vertical, rather than horizontal, temperature gradient \citep{Chauvet2012}. The similarity of the two problems is not coincidental: due to the weakness of advective fluxes, the {transport of heat and mass in the gas layer is primarily due to diffusion of the perturbations about the base state. This diffusive transport is independent of the direction of thermal gradient, which mainly affects the base state.}

Even when phase change is suppressed (e.g., at atmospheric conditions or for a fluid with relatively low volatility), we find that the gas phase can have a noticeable effect {on the stability of the flow}. Our analysis shows that the transport of heat in the gas layer can account for an increase {of more than ten} percent in the critical Marangoni number. The effect of heat transport through the gas layer has been traditionally described using Newton cooling law, with the magnitude of the heat flux characterized by the Biot number. While this {law} may be a reasonable approximation for the base flow, it breaks down completely the moment convection rolls appear. {In fact, heat transport in the gas phase has to be treated explicitly in order to obtain quantitatively accurate results in the entire range of parameters.}

\section*{Acknowledgements}
This work was supported in part by the National Science Foundation under Grant No. CMMI-1511470. 

\appendix

\section{Numerical Solution of the Boundary Value Problem\label{sec:numbvp}} 

{The boundary value solver {\bf bvp5c} is designed to deal with systems of first-order ODEs, so the higher-order ODEs arising in the linear stability analysis were converted to this form. }

\subsection{Diffusion-Dominated Case}

For the diffusion-dominated case, we converted the boundary value problem (\ref{eq:perl}) to a system of six first-order ODEs
\begin{align}
y_1'&=y_2,\qquad
y_2'=y_3,\nonumber \\
y_3'&=y_4,\qquad
y_4' =C_4(\tilde{z}) y_1 + C_5(\tilde{z}) y_3 + i Gr q y_5,\nonumber\\
y_5'& = y_6,\qquad
y_6' = C_6(\tilde{z}) y_1+Pr y_2+C_7(\tilde{z}) y_5,\label{eq:bvp}
\end{align}
where $y_1 = \tilde{\psi}_{lq}$, $y_5 = \tilde{\theta}_{lq}$, and
\begin{align}
C_4(\tilde{z})&=-i\,Gr \frac{q^3(\tilde{z}+1)(8{\tilde{z}}^2+\tilde{z}-1)}{48}
- i\,Gr\,q \frac{8\tilde{z}+3}{8}\nonumber \\
&+ i\,Re \frac{q^3(\tilde{z}+1)(3\tilde{z}+1)+6q}{4}
- q^2(q^2+\sigma_q),\nonumber  \\
C_5(\tilde{z})&= i\,Gr\frac{q (\tilde{z}+1)(8{\tilde{z}}^2+\tilde{z}-1)}{48}
-i\,Re\frac{q (\tilde{z}+1)(3\tilde{z}+1)}{4}+2q^2+\sigma_q,\nonumber \\
C_6(\tilde{z})&= i\,Ma\, Pr\, q\,C_3(\tilde{z}),\nonumber \\
C_7(\tilde{z}) &= -i\,Ma\,q\, C_1(\tilde{z})+\sigma_q Pr+q^2.\label{eq:bvpcoef}
\end{align}
Respectively, the boundary conditions become
\begin{align}
&y_1(-1)=0,\qquad
y_1(0)=0,\nonumber \\
&y_2(-1) =0,\qquad
y_6(-1) = 0,\nonumber \\
&y_3(0) + iq\, Re\, y_5(0)=0
\label{eq:bvpbc}
\end{align}
and
\begin{align}\label{eq:bvpbc2}
y_6(0) + Bi_q y_5(0)=0.
\end{align}

\subsection{Transient Dynamics in the Gas Layer}

To avoid solving coupled boundary value problems \eqref{eq:perl} and \eqref{eq:per3} defined on adjacent domains, we remapped the domain of $\tilde{\theta}_{gq}$ and $\tilde{\varsigma}_{vq}$ from $[0,A]$ to $[-1,0]$ by defining $\bar{z}=-\tilde{z}/A$ and introducing two new functions $\bar{\theta}_{gq}(\bar{z})=\tilde{\theta}_{gq}(-\tilde{z})$ and $\bar{\varsigma}_{vq}(\bar{z})=\tilde{\varsigma}_{vq}(-\tilde{z})$. In terms of these new functions, the system (\ref{eq:per2}) becomes
\begin{align}
A^{-2}\bar{\theta}_{gq}''-q^2\bar{\theta}_{gq}
&=\sigma_q K_\alpha^{-1}\bar{\theta}_{gq},\nonumber\\
\label{eq:geqg8}
A^{-2}\bar{\varsigma}_{vq}''-q^2\bar{\varsigma}_{vq}
&=\sigma_q K_D^{-1}\bar{\varsigma}_{vq},
\end{align}
where the prime now denotes the derivative with respect to $\bar{z}$, and
the boundary condition \eqref{eq:qper3} is replaced with
\begin{align}
\label{eq:qper2a}
A\tilde{\theta}_{lq}'(0)&= -\frac{k_g}{k_l}\bar{\theta}'_{gq}(0)
- \frac{1-c_a^0}{c_a^0}H \bar{\varsigma}'_{vq}(0).
\end{align}
Furthermore, the no-flux boundary conditions at the top of the gas layer become
\begin{align}
\bar{\theta}'_{gq}(-1) &= 0,\nonumber\\
\bar{c}'_{vq}(-1) &= 0. \label{eq:bcngb}
\end{align}
Temperature continuity and the local phase equilibrium (\ref{eq:rhoviq}) at the free surface yield
\begin{align}
\bar{\theta}_{gq}(0)&=\tilde{\theta}_{lq}(0),\nonumber\\
\bar{\varsigma}_{vq}(0)&=\tilde{\theta}_{gq}(0). \label{eq:bcngt}
\end{align}

Converting (\ref{eq:geqg8}) into a system of first-order ODEs yields four additional equations
\begin{align}
y_7'& = y_8,\qquad\
y_8'= A^2(q^2+K_\alpha^{-1}\sigma_q)y_7,\nonumber\\
y_9'& = y_{10},\qquad
y_{10}'= A^2(q^2+K_D^{-1}\sigma_q)y_9,\label{eq:bvpg1}
\end{align}
where $y_7=\bar{\theta}_{gq}$, and $y_9=\bar{\varsigma}_{vq}$, which should be added to the system (\ref{eq:bvp}). The boundary conditions (\ref{eq:bvpbc}) remain and \eqref{eq:bvpbc2} is replaced with
\begin{align}
&Ay_6(0) +\frac{k_g}{k_l} y_8(0)+\frac{1-c_a^0}{c_a^0}H y_{10}(0)=0,\nonumber\\
&y_7(0) =y_5(0),\qquad
y_8(-1) =0,\nonumber\\
&y_9(0) =y_5(0),\qquad
y_{10}(-1) =0.\label{eq:bvpbcg2}
\end{align}

\subsection{The Effect of Advection in the Gas Layer}

Just like in the previous case, we introduce new functions $\bar{\theta}_{gq}(\bar{z})=\tilde{\theta}_{gq}(-\tilde{z})$, $\bar{\varsigma}_{vq}(\bar{z})=\tilde{\varsigma}_{vq}(-\tilde{z})$, and $\bar{\psi}_{gq}(\bar{z})=\tilde{\psi}_{gq}(-\tilde{z})$, so the system (\ref{eq:per3}) becomes
\begin{align}
{K_\nu}\bar{\nabla}_q^2\bar{\nabla}_q^2\bar{\psi}_{gq} + iq \bar{C}_1(\bar{z})\mathcal{R} A^{-2}\bar{\psi}''_{gq}  - iq \bar{C}_2(\bar{z}) \mathcal{R} \bar{\psi}_{gq} - iq \Xi_T \bar{\theta}_{gq} - iq\Xi_\varsigma \bar{\varsigma}_{vq} &  =\sigma_q\bar{\nabla}_q^2\bar{\psi}_{gq},\nonumber\\
K_\alpha\bar{\nabla}_q^2\bar{\theta}_{gq} + iq\bar{C}_1(\bar{z})\mathcal{R}\bar{\theta}_{gq} - iq\bar{C}_3(\bar{z}) \mathcal{R} K_\alpha^{-1}\bar{\psi}_{gq} + A^{-1}\bar{\psi}'_{gq}&= \sigma_q\bar{\theta}_{gq},
\nonumber\\
\label{eq:per3a}
K_D\bar{\nabla}_q^2\bar{\varsigma}_{vq} + iq\bar{C}_1(\bar{z})\mathcal{R}\bar{\varsigma}_{vq} - iq\bar{C}_3(\bar{z}) \mathcal{R} K_D^{-1}\bar{\psi}_{gq} + A^{-1}\bar{\psi}'_{gq}&= \sigma_q\bar{\varsigma}_{vq}.
\end{align}
where we have defined $\bar{\nabla}_q^2=A^{-2}\partial_{\bar{z}}^2-q^2$ and
\begin{align}
\bar{C}_1(\bar{z}) &=\frac{(\bar{z}+1)(3\bar{z}+1)}{4} 
- \mathcal{B}\frac{(\bar{z}+1)(8\bar{z}^2+\bar{z}-1)}{48},\nonumber\\
\bar{C}_2(\bar{z}) &=q^2\bar{C}_1(\bar{z})+\frac{3}{2A^2}
- \mathcal{B}\frac{8\bar{z}+3}{8A^2},\nonumber\\
\bar{C}_3(\bar{z}) &=-\frac{A\bar{z}(\bar{z}+1)^2}{4} 
+ \mathcal{B}\frac{A\bar{z}(\bar{z}+1)^2(2\bar{z}-1)}{48}.
\end{align}
The functions $\bar{\varsigma}_{vq}(\bar{z})$ and $\bar{\theta}_{gq}(\bar{z})$ again satisfy the boundary conditions (\ref{eq:bcngb}) and (\ref{eq:bcngt}). The boundary conditions for the function $\bar{\psi}_{gq}(\bar{z})$ follow from (\ref{eq:bcvgb}) and (\ref{eq:bcvgt}):
\begin{align}
\bar{\psi}_{gq}(-1)&=0,\qquad
\bar{\psi}'_{gq}(-1)=0\nonumber\\
\bar{\psi}_{gq}(0)&=0,\qquad
\bar{\psi}'_{gq}(0)=-A\bar{\psi}'_{lq}(0).
\end{align}

Converting (\ref{eq:per3a}) into a system of first-order ODEs yields eight additional equations
\begin{align}
y_7'& = y_8,\qquad\ \
y_8'= \bar{C}_4(\bar{z})y_7+iq A^2 \mathcal{R} K_\alpha^{-2}\bar{C}_3(\bar{z})y_{11}-AK_\alpha^{-1}y_{12},\nonumber\\
y_9'& = y_{10},\qquad
y_{10}'= \bar{C}_5(\bar{z})y_9+iq A^2 \mathcal{R} K_D^{-2}\bar{C}_3(\bar{z})y_{11}-AK_D^{-1}y_{12},\nonumber \\
y_{11}'&= y_{12},\qquad
y_{12}'= y_{13},\nonumber \\
y_{13}'&= y_{14},\qquad
y_{14}'= iq A^4 K_\nu^{-1}(\Xi_T y_7+\Xi_\varsigma y_9)+\bar{C}_6y_{11}+\bar{C}_7y_{13},\label{eq:bvpg2}
\end{align}
which augment the system (\ref{eq:bvp}) describing the liquid layer. Here $y_7=\bar{\theta}_{gq}$, $y_9=\bar{\varsigma}_{vq}$,  $y_{11}=\bar{\psi}_{gq}$, and
\begin{align}
\bar{C}_4(\bar{z})&=-iq A^2 \mathcal{R} K_\alpha^{-1}\bar{C}_1(\bar{z}) + A^2(q^2+K_\alpha^{-1}\sigma_q),\nonumber  \\
\bar{C}_5(\bar{z})&= -iq A^2 \mathcal{R} K_D^{-1}\bar{C}_1(\bar{z}) + A^2(q^2+K_D^{-1}\sigma_q),\nonumber \\
\bar{C}_6(\bar{z})&=iq A^4 \mathcal{R} K_\nu^{-1}\bar{C}_2(\bar{z}) - A^4q^2(q^2+K_\nu^{-1}\sigma_q),\nonumber  \\
\bar{C}_7(\bar{z})&=-iq A^2 \mathcal{R} K_\nu^{-1}\bar{C}_1(\bar{z}) + A^2(2q^2+K_\nu^{-1}\sigma_q).\label{eq:bvpgcoef}
\end{align}
The boundary conditions are given by \eqref{eq:bvpbc}, \eqref{eq:bvpbcg2}, and
\begin{align}
y_{11}(0) &=0,\hspace{19.5mm}
y_{11}(-1) =0,\nonumber\\
y_{12}(0) &=-A y_2(0),\qquad
y_{12}(-1) =0.
\label{eq:bvpbcg1}
\end{align}
In all of the expressions given in the Appendix, the nondimensional parameters $Re$, $Gr$, $\mathcal{R}$, $\mathcal{B}$, $\Xi_T$, and $\Xi_\varsigma$ were expressed in terms of $Ma$, $Pr$, $Bo_D$, and $c_a^0$ using the definitions \eqref{eq:BoD}, \eqref{eq:Gr}, \eqref{eq:Xi}, \eqref{eq:Re}, \eqref{eq:Omega}, \eqref{eq:RB}, and \eqref{eq:Xi2}. Similarly, parameters $K_D$, $K_\alpha$ and $K_\nu$ were evaluated as functions of $c_a^0$.


\bibliographystyle{jfm}
\bibliography{../bibtex/heatpipe}

\end{document}